\documentclass[a4paper,fleqn,usenatbib]{mnras}
\usepackage{subfig}
\usepackage[T1]{fontenc}
\usepackage{ae,aecompl}
\usepackage{graphicx} 
\usepackage{amsmath}  
\usepackage{amssymb}    
\usepackage{lscape}
\usepackage{rotating}

\usepackage{xcolor} 
\usepackage{soul} 

\begin{document}
\title{The X-ray Ribs Within the Cocoon Shock of Cygnus A}
\author[R.~T.~Duffy et al.]
{\parbox{\textwidth}{R.~T.~Duffy$^{1}$\thanks{E-mail: \texttt{r.duffy@bristol.ac.uk}},
D.~M.~Worrall$^{1}$,
M.~Birkinshaw$^{1}$,
P.~E.~J.~Nulsen$^{2,3}$,
M.~W.~Wise$^{4,5}$,
M.~N.~de~Vries$^{5}$,
B.~Snios$^{2}$,
W.~G.~Mathews$^{6}$,
R.~A.~Perley$^{7}$,
M.~J.~Hardcastle$^{8}$,
D.~A.~Rafferty$^{9}$,
B.~R.~McNamara$^{10}$,
A.~C.~Edge$^{11}$,
J.~P.~McKean$^{4,12}$,
C.~L.~Carilli$^{7,13}$,
J.~H.~Croston$^{14}$,
L.~E.~H.~Godfrey$^{4}$
and R.~A.~Laing$^{15}$}\vspace{0.4cm}\\
\parbox{\textwidth}{$^1$HH Wills Physics Laboratory, University of Bristol, Tyndall Avenue, Bristol, BS8 1TL, UK\\
$^2$Harvard-Smithsonian Center for Astrophysics, 60 Garden Street, Cambridge MA 02138, USA\\
$^3$ICRAR, University of Western Australia, 35 Stirling Hwy, Crawley, WA 6009, Australia\\
$^4$ASTRON (Netherlands Institute for Radio Astronomy), PO Box 2, 7990 AA Dwingeloo, the Netherlands\\
$^5$Astronomical Institute `Anton Pannekoek', University of Amsterdam, Postbus 94249, 1090 GE Amsterdam, the Netherlands\\
$^6$University of California Observatories/Lick Observatory, Department of Astronomy and Astrophysics, University of California, Santa Cruz, CA 95064, USA\\
$^7$National Radio Astronomy Observatory, P.O. Box O, Socorro, NM 87801, USA\\
$^8$Centre for Astrophyics Research, School of Physics, Astronomy and Mathematics, University of Hertfordshire, College Lane, Hatfield AL10 9AB, UK\\
$^9$Hamburger Sternwarte, Universitat Hamburg, Gojenbergsweg 112, 21029, Hamburg, Germany\\
$^{10}$Department of Physics and Astronomy, University of Waterloo, 200 University Avenue West, Waterloo, ON N2L 3G1, Canada\\
$^{11}$Centre for Extragalactic Astronomy, University of Durham, Durham DH1 3LE, UK\\
$^{12}$Kapteyn Astronomical Institute, PO Box 800, NL-9700 AV Groningen, the Netherlands\\
$^{13}$Astrophysics Group, Cavendish Laboratory, JJ Thomson Avenue, Cambridge CB3 0HE, UK \\
$^{14}$School of Physical Sciences, The Open University, Walton Hall, Milton Keynes, MK7 6AA, UK\\
$^{15}$Square Kilometre Array Organisation, Jodrell Bank Observatory, Lower Withington, Macclesfield, Cheshire SK11 9DL, UK}}
\maketitle

\begin{abstract}
We use new and archival \emph{Chandra} observations of Cygnus A, totalling $\sim$1.9 Ms, to investigate the distribution and temperature structure of gas lying within the projected extent of the cocoon shock and exhibiting a rib-like structure. We confirm that the X-rays are dominated by thermal emission with an average temperature of around 4 keV, and have discovered an asymmetry in the temperature gradient, with the southwestern part of the gas cooler than the rest by up to 2 keV. Pressure estimates suggest that the gas is a coherent structure of single origin located inside the cocoon, with a mass of roughly $2\times10^{10} M_{\odot}$. We conclude that the gas is debris resulting from disintegration of the cool core of the Cygnus A cluster after the passage of the jet during the early stages of the current epoch of activity. The 4 keV gas now lies on the central inside surface of the hotter cocoon rim. The temperature gradient could result from an offset between the centre of the cluster core and the Cygnus A host galaxy at the switch-on of current radio activity.
\end{abstract}


\section{Introduction}
Cygnus A is the best known and nearest example of a powerful FRII radio galaxy. The galaxy which hosts Cygnus A is also the dominant galaxy of a 6 keV cluster with mass within 500 kpc in excess of $\sim2\times10^{14} M_{\odot}$ \citep{1984Arnaud,2002Smith}. The proximity and power of Cygnus A allows in-depth study of interactions between the expanding radio lobes of Cygnus A and the surrounding X-ray gas.

The host galaxy of Cygnus A resides within an elliptical cocoon shock, which extends about 60 arcsec (65 kpc) from north to south of the AGN and about 120 arcsec (130 kpc) from west to east \citep{2015Nulsen,2017Snios}. Within the projected extent of the cocoon shock, surrounding the radio core, lies a region of enhanced X-ray emission with a complex rib-like structure. Previous studies of this have revealed it to be thermal emission from gas with an average temperature of about 4 keV \citep{2002Smith, 2006Wilson}. The X-ray emitting ribs lie predominantly between the high-frequency radio lobes, their long axes mostly perpendicular to the radio axis of Cygnus A. We call this structure `rib-like' as its appearance suggests a hollow structure, with the filaments lying in the surrounding envelope, so that they are seen projected on to the core from in front or behind, rather than lying within it. It is conceivable that the origin of the gas may be similar to that of belt-like emission seen in other local radio galaxies \citep{2007Worrall,2007Hardcastle,2013Mannering,2016Duffy}, and enhanced central gas distributions showing interactions with the base of the lobes are a common feature of numerical models of radio galaxies \citep{2013Hardcastle,2014Hardcastle}.

Despite spectral studies of the gas, no firm conclusions have been drawn as to its origin. One possibility suggested by \citet{2002Smith} is that the rib-like structure represents a large-scale accretion disk with gas flowing into the Cygnus A nucleus. Alternatively, \citet{2010Mathews} suggest that the X-ray emission may be the expanded remnant of denser gas formerly located near the centre of Cygnus A, which has been shocked and heated by the AGN during the early stages of FRII development. 
 
Using archival and new \emph{Chandra} data, totalling over 1.9 Ms of observation, we investigate the physical properties of the gas interior to the cocoon shock of Cygnus A, specifically focussing on the rib-like structure which surrounds the core. Throughout this paper a flat $\Lambda$CDM cosmology with $\Omega_{m_{0}}$=0.3 and $\Omega_{\Lambda_{0}}$=0.7 with $H_{0}$=70 km s$^{-1}$ Mpc$^{-1}$ is adopted. This gives an angular scale of 1.088 kpc/arcsec at $z$=0.056075 \citep{1997Owen}, the redshift of Cygnus A.

\section{Chandra Observations}
Cygnus A has been observed numerous times using \emph{Chandra's} Advanced CCD Imaging Spectrometer (ACIS), with the source centred on either the S3 or I3 chip, mostly as part of the Chandra Visionary Project \citep{2017Wise}. We have opted to include only observations where Cygnus A is centred on the I3 chip, as these form the vast majority of observations. We did not include any observations where Cygnus A's northwest subcluster was the target, or ObsIDs 17145 and 17505 where the PSF near Cygnus A's core is very broad. A list of the \emph{Chandra} observations used in the present work is given in Table \ref{table:obs}. 

\begin{table}
\centering
	\begin{tabular}{ccc|ccc}
	\hline
	ObsID & Date & $t_{exp}$ & ObsID & Date & $t_{exp}$ \\
	\hline
	05830 & 2005-05-26 & 23.5 & 17518 & 2016-07-16 & 49.4 \\
	05831 & 2005-02-16 & 51.1 & 17519 & 2016-12-19 & 29.7 \\
	06225 & 2005-02-15 & 24.3 & 17520 & 2016-12-06 & 26.8 \\
	06226 & 2005-02-19 & 23.8 & 17521 & 2016-07-20 & 24.7 \\
	06228 & 2005-02-25 & 16.0 & 17522 & 2017-04-08 & 49.4 \\
	06229 & 2005-02-23 & 22.8 & 17523 & 2016-08-31 & 49.4 \\
	06250 & 2005-02-21 & 7.0 & 17524 & 2015-09-08 & 23.0 \\
	06252 & 2005-09-07 & 29.7 & 17525 & 2017-04-22 & 24.7 \\
	17133 & 2016-06-18 & 30.2 & 17526 & 2015-09-20 & 49.4 \\
	17134 & 2017-05-20 & 29.4 & 17527 & 2015-10-11 & 26.7 \\
	17135 & 2017-01-20 & 19.8 & 17528 & 2015-08-30 & 49.3 \\
	17136 & 2017-01-26 & 22.2 & 17529 & 2016-12-15 & 35.1 \\
	17137 & 2017-03-29 & 25.2 & 17530 & 2015-04-19 & 21.3 \\
	17138 & 2016-07-25 & 26.4 & 17650 & 2015-04-22 & 28.2 \\
	17139 & 2016-09-16 & 39.5 & 17710 & 2015-08-07 & 19.8 \\
	17140 & 2016-10-02 & 34.6 & 18441 & 2015-09-14 & 24.6 \\
	17141 & 2015-08-01 & 29.7 & 18641 & 2015-10-15 & 22.4 \\
	17142 & 2017-04-20 & 23.3 & 18682 & 2015-10-14 & 22.8 \\
	17143 & 2015-09-04 & 27.1 & 18683 & 2015-10-18 & 14.1 \\
	17144 & 2015-05-03 & 49.4 & 18688 & 2015-11-01 & 34.6 \\
	17507 & 2016-11-12 & 32.6 & 18871 & 2016-06-13 & 21.8 \\
	17508 & 2015-10-28 & 14.8 & 18886 & 2016-07-23 & 21.7 \\
	17509 & 2016-07-10 & 51.4 & 19888 & 2016-10-01 & 19.5 \\
	17510 & 2016-06-26 & 37.5 & 19956 & 2016-12-10 & 54.3 \\
	17511 & 2017-05-11 & 15.9 & 19989 & 2017-02-12 & 41.5 \\
	17512 & 2016-09-25 & 66.8 & 19996 & 2017-01-28 & 28.9 \\
	17513 & 2016-08-15 & 49.4 & 20043 & 2017-03-25 & 29.6 \\
	17514 & 2016-12-13 & 49.4 & 20044 & 2017-03-27 & 14.8 \\
	17515 & 2017-03-22 & 39.5 & 20048 & 2017-05-19 & 22.8 \\
	17516 & 2016-08-18 & 49.0 & 20077 & 2017-05-13 & 27.7 \\
	17517 & 2016-09-17 & 26.7 & 20079 & 2017-05-21 & 23.8 \\
	\hline
	& & & & Total: & 1919.8 ks \\		
	\hline
	\end{tabular}
	\caption{List of \emph{Chandra} observations used in this work, with observation identification number, start date of the observation and net exposure time after background flare corrections, $t_{exp}$, in ks.}
	\label{table:obs}
\end{table}

The data were reprocessed using \textsc{chandra\_repro} from \texttt{CIAO} 4.8 or 4.9 with CALDB 4.7.2 or 4.7.4 (later versions used for 2017 observations) and were then cleaned using the routine \textsc{deflare} in its \emph{lc\_clean} mode to remove the small contribution from background flaring, with applied GTIs removing no more than 1 ks of exposure time in any observation. To account for astrometric errors ObsID 5831 was chosen as a reference due to its high total counts. Following \citet{2017Snios}, 0.5 - 7.0 keV images in 0.492-arcsec pixels were made of a region 160 arcsec by 120 arcsec centred on Cygnus A, with the events from other ObsIDs reprojected into the same sky frame as ObsID 5831. The cross correlations between each ObsID's image and the image from ObsID 5831 were then fitted with a Lorentzian profile to determine the offsets between them. These offsets of order 0.5 arcsec or less constitute the astrometric shifts required to align the data sets with ObsID 5831. The offsets were applied to the event and aspect solution files using the \textsc{wcs\_update} command in \texttt{CIAO}. Aligned exposure corrected images were then merged together using \textsc{merge\_obs} for image analysis.

\section{X-ray Morphology}
\begin{figure*}
	\centering
	\includegraphics[width=0.99\textwidth]{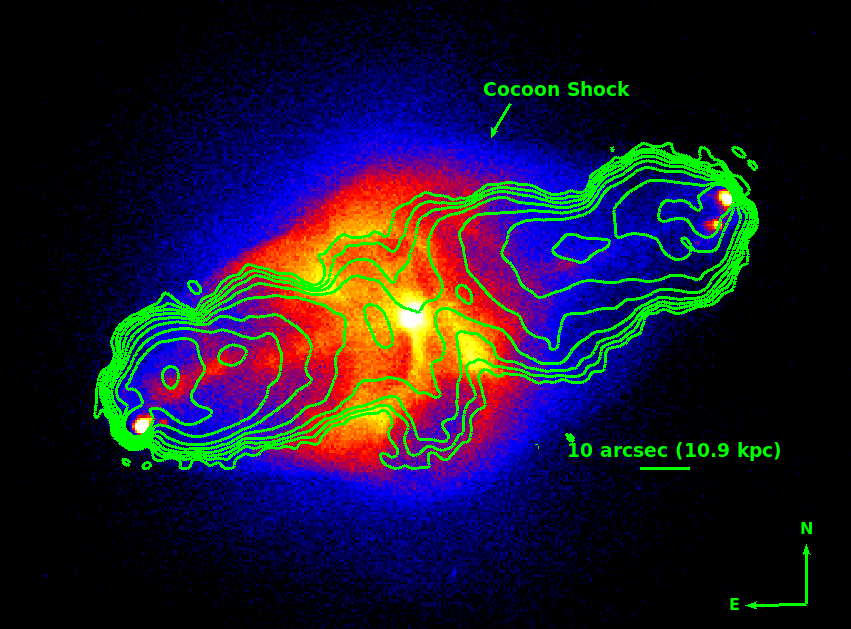}
	\caption{Merged \emph{Chandra} exposure-corrected 0.5-7.0 keV image of Cygnus A observations listed in Table \ref{table:obs}. with native 0.492 arcsec pixels. Contours are of a 327 MHz VLA radio map with restoring beam of size 2.75 arcsec from program AK570, and are shown at brightness levels of 1.434$\times$(0.2, 0.5, 1.0, 2.0, 5.0, 7.5, 15, 20, 25) Jy beam$^{-1}$.}
	\label{fig:xrayobs}
\end{figure*}

Figure \ref{fig:xrayobs} shows an exposure-corrected image made using all the observations listed in Table \ref{table:obs}. The edges of the cocoon shock are clearly visible at the lower brightness limit of the scale selected, as well as hotspots at the extremities of the lobes. Visible in both lobes is a broad X-ray jet-like feature, which is roughly aligned with the radio jet axis close to the AGN, but becomes misaligned in the outer part of the jet \citep{2008Steenbrugge}. The bright eastern X-ray jet-like feature can be seen extending almost the full length towards the eastern hot spots, while the western X-ray jet-like feature can only be seen extending roughly half way along the length of the western lobe in this particular image. The core can be seen between the lobes at the centre of the bright rib-like emission. The ribs occupy a region between high-brightness parts of the radio lobes, although radio emission extends across the centre of the lobes and down to the south of the ribs and is more uniform at lower radio frequencies. There is no obvious relation between the X-ray brightness enhancements associated with the ribs and the radio emission. The brightest part of the X-ray rib structure is seen extending immediately to the south of the core, with another bright filament to the west of this at the boundary of the ribs. The north-south axis of the ribs measures roughly 42 arcsec (45 kpc) in length, while the east-west axis measures roughly 29 arcsec (32 kpc). To the south of the brightest rib structure and contours on Figure \ref{fig:xrayobs} is a roughly elliptical region of low surface brightness representing the cavity previously studied by \citet{2012Chon}.

\subsection{Unsharp Mask Images}

\begin{figure}
	\centering
	\includegraphics[width=0.99\columnwidth]{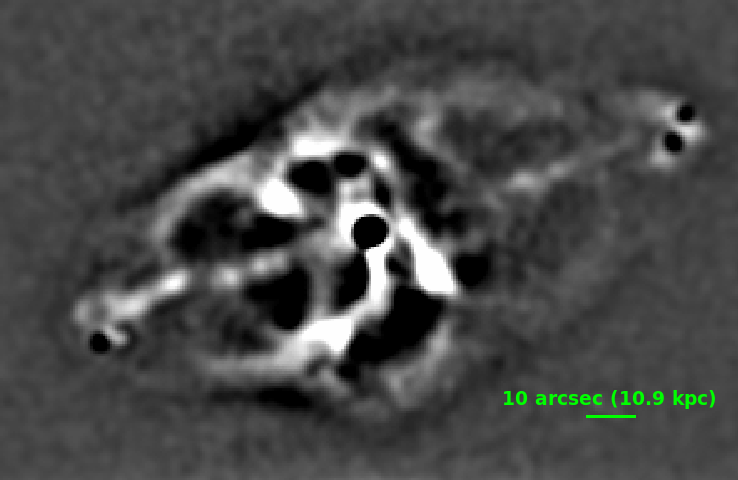}
	\includegraphics[width=0.99\columnwidth]{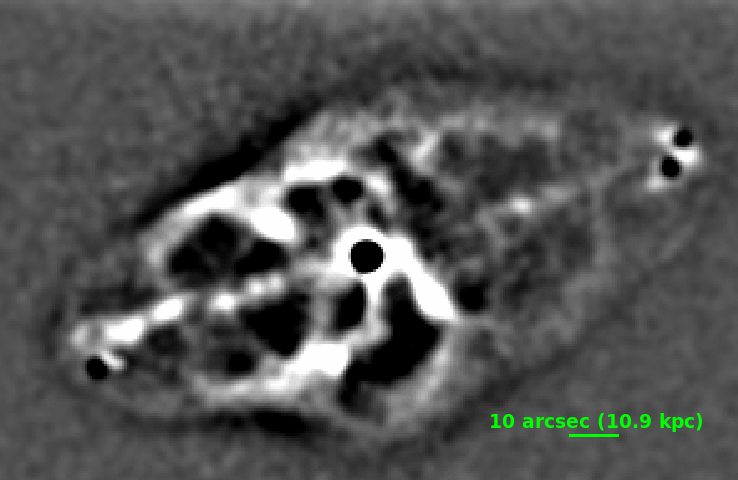}
	\caption{\emph{Top:} Unsharp mask image made from the soft 0.5-2.0 keV band by subtracting an image smoothed with a Gaussian dispersion of 10 arcsec from one smoothed by 2 arcsec. \emph{Bottom:} Unsharp mask image made from the hard 2.0-7.0 keV band by subtracting an image smoothed with a Gaussian dispersion of 10 arcsec from one smoothed by 2 arcsec. Black circular regions correspond to excised core and hotspots.}
	\label{fig:unsharpmask}
\end{figure}

The merged X-ray image was used to create unsharp mask images of Cygnus A. Firstly the image was binned into soft (0.5 - 2.0 keV) and hard (2.0 - 7.0 keV) energy bands. The core to a radius of 3 arcsec and the hotspots were excised. The hard and soft images were then smoothed on two different length scales and these images were subtracted to produce the unsharp mask images. Figure \ref{fig:unsharpmask} shows the result after using Gaussian smoothings of 2 arcsec and 10 arcsec.

In both panels of Figure \ref{fig:unsharpmask} the rib structure is clearly visible. The rib to the south is unusual in that it appears to terminate at the AGN, unlike any of the others, although this is possibly an effect of projection. There appear to be several holes, corresponding to regions of low surface brightness, not only visible within the rib structure, but also extending beyond the position where the south rib bends. The unsharp mask images also clearly show X-ray jet-like features of Cygnus A to both the east and the west. 

\section{Spectroscopy}
\subsection{Region Selection}
The structure of the X-ray emission from the region lying between the bright radio lobes is complex and clearly inhomogeneous. We first divided it into related regions for spectral analysis by applying the \textsc{contbin 1.4} software \citep{2006Sanders} on the merged data. \textsc{Contbin} employs an algorithm for spatially binning X-ray data using contours on an adaptively smoothed map and defining regions of matched signal-to-noise ratio (SNR). The regions closely match the surface brightness structure, allowing us to examine whether different populations of gas are found in different ribs of Cygnus A. 

We tested \textsc{contbin} with several different settings, before settling on a SNR of 200 after the image had been adaptively smoothed with a SNR of 15. The resulting regions aligned well with the surface brightness features identifiable by eye. The region within 5 arcsec of the AGN and a broad rectangular region around the eastern jet-like emission were excluded from all analysis. No background was used in defining the regions, as  the features we are studying are much brighter than their surroundings. \textsc{Contbin} identified 28 regions across a 60 arcsec by 63 arcsec field.

\begin{figure*}
	\centering
	\includegraphics[width=0.99\textwidth]{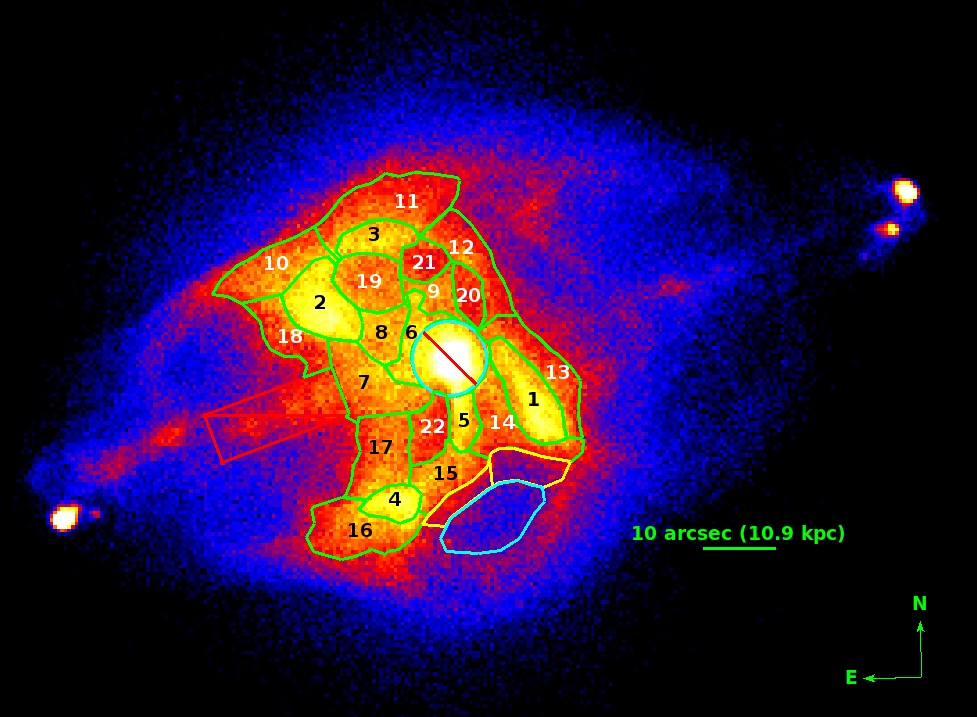}
	\caption{Map identifying spectral extraction regions across the rib-like structure of Cygnus A. The unnumbered cyan region to the south corresponds to the position of the \citet{2012Chon} cavity, while the yellow regions correspond to regions between the ribs and cavity. A rectangular region coincident with the jet-like feature and a 5 arcsec region are excluded and not used in spectral analysis.}
	\label{fig:regselfdefined}
\end{figure*}

Whilst the \textsc{contbin} regions were interesting in helping us to define appropriately sized and located regions for selection, this selection misses several potentially interesting structures. This applies mostly to regions of low surface brightness, which are amalgamated into other larger regions in order for \textsc{contbin}'s selection to reach the appropriate SNR. To prevent this, we adjusted the regions suggested by \textsc{contbin} to make the connected regions more contiguous and to create smoother boundaries with the regions of lower surface brightness while roughly retaining the signal to noise. We identified 22 regions on which to perform spectral extraction, across a smaller area than that used in the initial \textsc{contbin} selection. The 22 regions are shown in Figure \ref{fig:regselfdefined}.

\subsection{Spectral Fitting}
\label{sec:spectralfitting}
\begin{figure}
	\centering
	\includegraphics[width=0.99\columnwidth]{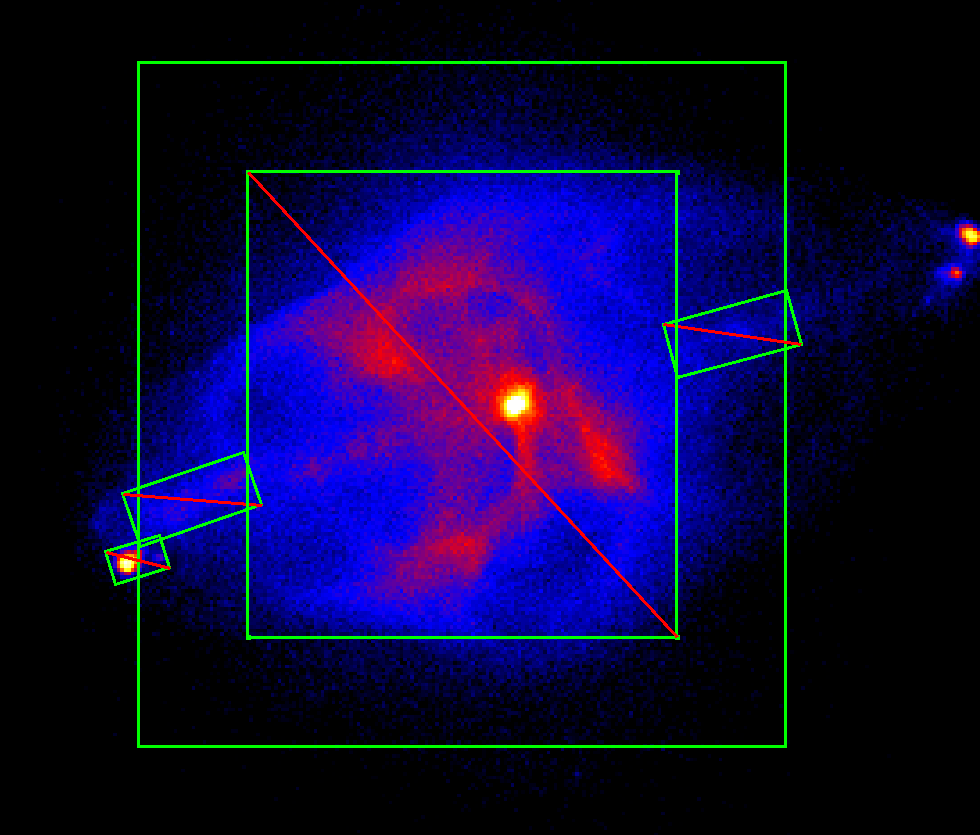}
	\caption{Merged \emph{Chandra} image of Cygnus A. The `square annulus' local background extraction region is overlaid.}
	\label{fig:localbkg_square}
\end{figure}

We utilised \textsc{specextract} to extract source and local background spectra for each of the 22 regions from the 50 included observations. The local background is taken from a square annulus outside our area of interest, with width 15 arcsec (see Figure \ref{fig:localbkg_square}). \texttt{CIAO}'s \textsc{combine\_spectra} was used to sum the spectra from individual observations and combine the calibration files, with the counts binned to retain at least 20 counts per bin to allow the use of $\chi^{2}$ fitting. We found that combining data with very different observation times had no significant impact on the fits, and results were consistent between early and late data. We applied \textsc{readout\_bkg} to each observation to estimate the contribution of out-of-time (OOT) events to each spectrum and remove it, although the rib region is sufficiently bright that the OOT contribution is negligible.

Each of the 22 regions was fitted in \texttt{XSPEC} \citep{1996Arnaud} with a PHABS$\times$APEC model \citep{2001Smith} between 0.5 and 7.0 keV with $z$=0.056075 and Galactic $N_{H}=3.1\times10^{21}$ cm$^{-2}$ based on an average of results from \citet{1990Dickey} and \citet{2005Kalberla}. The abundance was left free in the fitting and was scaled to the solar abundances of \citet{1989Anders}. Freeing the PHABS component of our model showed that $N_{H}=3.1\times10^{21}$ cm$^{-2}$ is an underestimate of the absorption in the region of the ribs. To account for absorption associated with the host galaxy of Cygnus A, we added a second absorption component to our model, ZPHABS. The ZPHABS component increases the value of $N_{H}$ in almost all regions by at least 10 per cent. The highest absorption is in the north ribs. A broadly acceptable fit could be found for each region using just this model, suggesting that each region is dominated by single-temperature thermal emission. 

The origin of excess photoelectric absorption in brightest cluster galaxies (BCG) is a topic of current debate, with suggested origins including gas condensing from lower entropy gas which is lifted outwards from the cluster core by X-ray bubbles \citep{2016McNamara} or gravitational attraction from the BCG moving through a region with a short cooling time focussing cooling in its wake \citep{2001Fabian}. The excess absorption we detect is similar to the column densities of molecular gas seen in other BCGs \citep{2016Vantyghem}. \citet{2000Wilman} detect molecular hydrogen extending some 6 kpc from the centre of Cygnus A, although the highest absorption columns we measure are found at around 20 kpc from the core.

We expect some systematic uncertainties in combining 62 data sets, but as a goal we have defined an acceptable fit from our $\chi^{2}$ fitting as one with a null hypothesis probability $\geq$0.01. The majority of the selected regions fit acceptably by these criteria while modelled as an absorbed APEC model (see Table \ref{table:sdspecfits}). It is possible that some of these regions are contaminated with non-thermal core AGN emission. In our extraction regions, the region near the AGN was excised with a circle of radius 5 arcsec. However, comparison of the PSF (for the ACIS-S image, ObsID 1707, which has a 0.4 s frame time to mitigate the effect of pileup) with a radial profile suggests that significant core emission may be found up to 8 arcsec from the core. We therefore included the spectral model of the nucleus from \citet{2002Young} (see their Table 2, Model 1) in our poorly fit regions close to the core. We included both the hard and soft model components, and their relative normalisations were fixed. We also used the absorption column suggested in \citet{2002Young}, although this is likely to have varied across the timescale of the various observations \citep{2015Reynolds}. The core model is necessary as a simple power law does not improve the fits, and the data show a preference for a spectrum that rises with increasing energy. The shape of the core model is very different from a single-component power law and its inclusion causes a significant improvement for several regions.

Adding the core component brought the fits of regions 1, 5, 6 and 13 to an acceptable level and improved the fit of region 14. F-test probabilities are $\leq10^{-7}$ for each region, showing that the addition of the core components is very strongly favoured. Poorly fitted regions 2 and 11 are too far from the core to justify the inclusion of this model. We tested for multi-temperature components in regions still giving a poor fit, but could find no evidence of a second component. Radio maps at 327 MHz, 1345 MHz and 146 MHz show Cygnus A is bright at low radio frequencies across the centre and much of the southern ribs \citep{2006Lazio,2010Steenbrugge,2016McKean} and inverse-Compton X-ray emission at some level is expected, so we tested for an extra single power-law component. We again found no evidence for this component, although regions 2 and 11 occupy the northern part of the ribs away from the brightest radio emission. Inverse-Compton emission has been plausibly detected in the lobes of Cygnus A in the past \citep{2010Hardcastle,2010Yaji}, although these detections are where the radio emission is brightest at GHz frequencies, to the east and west of the ribs. The poor fits in regions 2 and 11 are possibly explained by mixing of gases, but it seems the mixing is sufficiently complicated that the spectral fitting does not cope well.

The best fit APEC components are given in Table \ref{table:sdspecfits}. Regions marked with a `*' are those where additional components were added to the absorbed APEC model to improve the fits. 

\begin{table*}
\begin{tabular}{l c c c c c c c}
	\hline
	Region & Source $N_{H}$ ($10^{21}$ cm$^{-2}$) & $kT$ (keV) & $Z$ ($Z_{\odot}$) & $N_{\rm APEC}$ & $N_{\rm hard\,core}$ &  $\chi^{2}$/dof & Probability  \\
	\hline
	*Region 1 & $0.19^{+0.16}_{-0.15}$ & $3.22^{+0.10}_{-0.11}$ & $0.78\pm0.06$ & $5.23^{+0.15}_{-0.14}$ & $0.20\pm0.05$ & 510.3/439 & 0.01 \\
	Region 2 & $0.67^{+0.16}_{-0.15}$ & $3.94\pm0.10$ & $0.72\pm0.05$ & $5.92\pm0.13$ & - & 481.6/396 & 0.002 \\
	Region 3 & $0.95\pm0.20$ & $3.76^{+0.14}_{-0.13}$ & $0.77^{+0.08}_{-0.07}$ & $2.68\pm0.08$ & - & 443.8/440 & 0.44 \\
	Region 4 & $0.21^{+0.22}_{-0.21}$ & $3.50^{+0.15}_{-0.13}$ & $0.59\pm0.07$ & $2.29^{+0.08}_{-0.07}$ & - & 510.2/440 & 0.01 \\
	*Region 5 & $0.30^{+0.31}_{-0.29}$ & $2.61^{+0.14}_{-0.15}$ & $0.69^{+0.11}_{-0.10}$ & $1.84^{+0.12}_{-0.11}$ & $0.12\pm0.03$ & 505.5/439 & 0.02 \\
	*Region 6 & $0.55^{+0.29}_{-0.27}$ & $3.26^{+0.18}_{-0.19}$ & $0.81^{+0.11}_{-0.10}$ & $2.15^{+0.11}_{-0.10}$ & $0.19\pm0.04$ & 436.3/439 & 0.53 \\
	Region 7 & $0.42\pm0.16$ & $4.90\pm0.17$ & $0.78\pm0.06$ & $4.46\pm0.10$ & - & 568.6/440 & $3.3\times10^{-5}$ \\
	Region 8 & $0.65^{+0.23}_{-0.21}$ & $3.83^{+0.15}_{-0.14}$ & $0.80\pm0.08$ & $2.41\pm0.08$ & - & 485.5/440 & 0.07 \\
	Region 9 & $0.48^{+0.27}_{-0.26}$ & $3.69^{+0.17}_{-0.16}$ & $0.88\pm0.10$ & $1.62\pm0.06$ & - & 508.9/440 & 0.01 \\
	Region 10 & $0.61^{+0.19}_{-0.18}$ & $4.73\pm0.19$ & $0.76\pm0.07$ & $3.74\pm0.10$ & - & 477.7/440 & 0.10 \\ 
	Region 11 & $0.84^{+0.18}_{-0.17}$ & $4.23^{+0.13}_{-0.12}$ & $0.65\pm0.05$ & $4.89\pm0.11$ & - & 528.9/440 & 0.002 \\
	Region 12 & $0.51^{+0.21}_{-0.20}$ & $4.55\pm0.21$ & $0.74\pm0.07$ & $2.80\pm0.08$ & - & 497.1/440 & 0.03 \\
	*Region 13 & $0.59^{+0.25}_{-0.23}$ & $3.44^{+0.25}_{-0.17}$ & $0.61\pm0.07$ & $3.21^{+0.13}_{-0.12}$ & $0.21^{+0.05}_{-0.06}$ & 476.1/439 & 0.11 \\
	*Region 14 & $0.41^{+0.22}_{-0.21}$ & $3.02^{+0.14}_{-0.15}$ & $0.81\pm0.08$ & $3.47\pm0.13$ & $0.28\pm0.05$ & 529.2/439 & 0.002 \\
	Region 15 & $0.21\pm0.20$ & $3.31\pm0.09$ & $0.74\pm0.07$ & $2.86\pm0.09$ & - & 483.4/440 & 0.07 \\
	Region 16 & $0.16^{+0.18}_{-0.16}$ & $4.27^{+0.15}_{-0.13}$ & $0.64\pm0.06$ & $4.01\pm0.10$ & - & 478.7/440 & 0.10 \\
	Region 17 & $0.08^{+0.17}_{-0.08}$ & $4.44\pm0.17$ & $0.78^{+0.07}_{-0.06}$ & $3.65\pm0.09$ & - & 475.2/440 & 0.12 \\
	Region 18 & $0.64^{+0.26}_{-0.25}$ & $4.63^{+0.25}_{-0.24}$ & $0.75^{+0.09}_{-0.08}$ & $2.44^{+0.09}_{-0.08}$ & - & 473.9/440 & 0.13 \\
	Region 19 & $0.55^{+0.21}_{-0.20}$ & $4.09\pm0.14$ & $0.75\pm0.07$ & $3.27\pm0.09$ & - & 510.2/440 & 0.01 \\
	Region 20 & $0.47^{+0.33}_{-0.31}$ & $4.63\pm0.32$ & $0.97^{+0.14}_{-0.13}$ & $1.05\pm0.05$ & - & 433.4/440 & 0.58 \\
	Region 21 & $0.58^{+0.34}_{-0.33}$ & $4.22^{+0.28}_{-0.21}$ & $0.87\pm0.12$ & $1.21^{+0.06}_{-0.05}$ & - & 280.4/268 & 0.29 \\
	Region 22 & $0.21^{+0.26}_{-0.21}$ & $4.10^{+0.18}_{-0.17}$ & $0.81\pm0.10$ & $1.66\pm0.06$ & - & 471.2/440 & 0.15 \\
	\hline
	\end{tabular}
	\caption{Results from spectral fits to each region. Regions 1, 5, 6, 13 and 14 have additional components ($N_{\rm APEC}$, $N_{\rm hard\,core}$) representing the core emission from \citet{2002Young} to make their fits acceptable. Column 1 is the absorption column at the source, obtained from the ZPHABS component of our model. All normalisations have units $10^{10}$ cm$^{-5}$. Errors are given for a 90\% confidence limit.}
	\label{table:sdspecfits}
\end{table*}

Region 7 has a particularly poor fit to the APEC model, which is not improved to acceptable levels by the inclusion of core or power-law components. This is likely to be because there is emission associated with the X-ray jet-like feature within this region. A fit including this is beyond the scope of this paper (but see \citealt{2017deVries}). For this reason, region 7 is not included in any analysis beyond the temperature mapping in Section \ref{sec:tempmap}.

A $\chi^{2}$ test to compare the abundances with a constant (each with 1$\sigma$ errors as compared with the 90\% values given in Table \ref{table:sdspecfits}) finds an unacceptable fit with $\chi^{2}$/dof = 67.4/21. This suggests some changes in metallicity across the rib structure. There appear to be no systematic trends, although regions of higher and lower metallicity are clustered together. The range in metallicities is consistent with the abundances seen in new radial profiles of the cluster gas up to 20 arcseconds beyond the cocoon shock\citep{2017Snios}. A fit to the cavity identified in \citet{2012Chon}, immediately south of the ribs gives a temperature of $4.39^{+0.31}_{-0.26}$ keV and an abundance of $0.62^{+0.11}_{-0.10} Z_{\odot}$.

Regions 1, 2, 3 and 4 are similar to those defined in other studies of the rib-like structure and can be compared with them. The temperatures we find tend to match those found by \citet{2002Smith} and are between 0.3 and 0.8 keV cooler than those found by \citet{2006Wilson} and \citet{2012Chon}. This is likely because we have used local background to remove the hot cluster component in the foreground and background, whereas the studies with which our results are discrepant were conducted using blank-sky backgrounds.

The metallicities and temperatures in two regions in the gap between the ribs and the cavity (shown in yellow in Figure \ref{fig:regselfdefined}) are the same ($kT=3.90\pm0.20$ keV and $Z=0.66^{+0.15}_{-0.14} Z_{\odot}$),  and are consistent with the metallicity of the cavity and of region 16. It is therefore plausible that region 16 and the cavity are composed of gas forming a continuous structure.

The location of our local background is an important consideration. We opted to use the square annulus seen in Figure \ref{fig:localbkg_square} in order to border the area we were using to identify spectral extraction regions using \emph{contbin}. The disadvantage of using this region is that instead of completely isolating the spectrum of the rib structures, we are including some emission from the shocked cocoon gas. This could potentially affect both temperatures and normalisations of the fits. To test the impact of the shocked rims on our spectral fits, we used the \citet{2012Chon} cavity as a local background on some of the best defined regions in the ribs. We opted to use the cavity as it lies within the cocoon shock so should contain foreground and background emission from the shocked rims, is close to the ribs, but does not form a part of the rib structure. The resulting fits in regions 1, 2 and 5 show no change in temperature and a reduction in normalisation of a factor of less than 1.5. This has a very small impact on calculation of the physical parameters, with pressures dropping by (0.1 - 0.2) $\times$ 10$^{-10}$ Pa in the selected regions.

\section{Physical Parameters of the Regions}
\label{sec:physpar}
The regions of the ribs are well described by thermal models, which return the temperatures and abundances within each. The \texttt{XSPEC} normalisation of the APEC component, $N_{\rm XSPEC}$, can be related to the proton density, $n_{\rm p}$, by Equation \ref{eqn:dens} \citep{2006Worrall}, where $D_{L}$ is the luminosity distance in cm, $V$ is the volume in cm$^{3}$.

\begin{align}
\label{eqn:dens}
\centering
n_{\rm p} \approx \sqrt{\frac{10^{14} N_{\rm XSPEC} 4 \pi D_{\rm L}^{2}}{(1+z)^{2} 1.18 V}}
\end{align}

If the proton density is constant over the volume, the pressure can then be found by 

\begin{align}
\label{eqn:pressure}
\centering
P = 3.6\times10^{-10}n_{\rm p}kT \quad\text{Pa}
\end{align}

\noindent where $n_{\rm p}$ is in units of cm$^{-3}$ and $kT$ is in units of keV \citep{2012Worrall}.

Additionally, we also calculated the entropy index, $S$, in each region using Equation \ref{eqn:entropy} \citep{1999Ponman}.

\begin{align}
\label{eqn:entropy}
\centering
S = \frac{kT}{n_{\rm p}^{2/3}}
\end{align}

We calculated the volumes of the regions within the ribs in two different ways. Firstly, we assumed the regions were cylinders or spheres represented by their projected volumes. Whether a region is circular or spherical is dependent on its 2D appearance. For example, region 19 is considered spherical, while region 1 is considered cylindrical. For cylinders, the volume was calculated assuming the longest axis of the region corresponds to the height of the cylinder and half the length of the shortest axis corresponds to the radius. The results using this method are shown in Table \ref{table:sdphyspar}. Secondly, we assume the rib structures each fill a cylinder of radius 39 arcsec (the semi-minor axis of the cocoon), lying in the plane of the sky, whose axis lies coincident with the radio axis. The surface areas of our regions are then multiplied by a depth representing the line of sight contained within the cylinder. The results using this method are shown in Table \ref{table:sdphysparalt}.

\begin{table*}
	\textbf{Cylinders/spheres geometry} \\
	\begin{tabular}{l c c c c}
	\hline
	Region & $n_{p}$ ($10^{-1}$ cm$^{-3}$) & $P$ ($10^{-10}$ Pa) & $M (10^{9} M_{\odot})$ & $S$ (keV cm$^{2}$) \\
	\hline
	Region 1 & $1.40\pm0.02$ & $1.6\pm0.1$ & $1.8\pm0.1$ & $11.9\pm0.4$\\
	Region 2 & $1.41\pm0.01$ & $2.0\pm0.1$ & $2.0\pm0.1$ & $14.5\pm0.4$ \\
	Region 3 & $1.83\pm0.01$& $2.5\pm0.1$ & $0.7\pm0.1$ & $11.7^{+0.5}_{-0.4}$ \\
	Region 4 & $1.68\pm0.03$ & $2.1\pm0.1$ & $0.7\pm0.1$ & $11.5^{+0.5}_{-0.4}$ \\
	Region 5 & $1.83^{+0.06}_{-0.50}$ & $1.7\pm0.1$ & $0.5\pm0.1$ & $8.1\pm0.5$ \\
	Region 6 & $2.29^{+0.06}_{-0.05}$ & $2.7\pm0.2$ & $0.5\pm0.1$ & $8.7\pm0.5$ \\
	Region 8 & $1.54\pm0.03$ & $2.1\pm0.1$ & $0.8\pm0.1$ & $13.3\pm0.5$ \\
	Region 9 & $1.33\pm0.02$ & $1.8\pm0.1$ & $0.6\pm0.1$ & $14.1^{+0.7}_{-0.6}$ \\
	Region 10 & $1.34\pm0.02$ & $2.3\pm0.1$ & $1.3\pm0.1$ & $18.0\pm0.7$ \\
	Region 11 & $1.22\pm0.01$ & $1.9\pm0.1$ & $1.9\pm0.1$ & $17.2\pm0.5$ \\
	Region 12 & $1.61\pm0.02$ & $2.6\pm0.1$ & $0.8\pm0.1$ & $15.4\pm0.7$ \\ 
	Region 13 & $1.75^{+0.04}_{-0.03}$ & $2.2^{+0.2}_{-0.1}$ & $0.9\pm0.1$ & $11.0^{+0.8}_{-0.6}$ \\
	Region 14 & $1.64\pm0.03$ & $1.8\pm0.2$ & $1.0\pm0.1$ & $10.1\pm0.5$ \\
	Region 15 & $1.73\pm0.03$ & $2.1\pm0.1$ & $0.8\pm0.1$ & $10.7\pm0.3$ \\
	Region 16 & $1.16\pm0.01$ & $1.8\pm0.1$ & $1.7\pm0.1$ & $18.0\pm0.6$ \\
	Region 17 & $1.28\pm0.02$ & $2.0\pm0.1$ & $1.4\pm0.1$ & $17.5\pm0.7$ \\
	Region 18 & $1.18\pm0.02$ & $2.0\pm0.1$ & $1.0\pm0.1$ & $19.2\pm1.0$ \\
	Region 19 & $1.15\pm0.02$ & $1.7\pm0.1$ & $1.4\pm0.2$ & $17.3\pm0.6$ \\
	Region 20 & $1.23\pm0.03$ & $2.0\pm0.2$ & $0.4\pm0.1$ & $18.7\pm1.3$ \\
	Region 21 & $1.21\pm0.03$ & $1.8\pm0.1$ & $0.5\pm0.1$ & $17.2^{+1.2}_{-0.9}$ \\
	Region 22 & $1.23\pm0.02$ & $1.8\pm0.1$ & $0.7\pm0.1$ & $16.6^{+0.8}_{-0.7}$ \\
	& & \textbf{Total mass:} & $21.0\pm1.0$ \\
	\hline
	\end{tabular}
	\caption{Density, pressure, mass and entropy index of the X-ray emission in the various regions within the rib structure of Cygnus A. Errors are quoted for 90\% confidence range. These were calculated assuming a  spherical or cylindrical volume for each region.}
	\label{table:sdphyspar}
\end{table*}

\begin{table*}
	\textbf{Filled cylinder geometry} \\
	\begin{tabular}{l c c c c}
	\hline
	Region & $n_{p}$ ($10^{-2}$ cm$^{-3}$) & $P$ ($10^{-11}$ Pa) & $M (10^{9} M_{\odot})$ & $S$ (keV cm$^{2}$)  \\
	\hline
	Region 1 & $3.73\pm0.05$ & $4.3^{+0.1}_{-0.2}$ & $6.7\pm0.1$ & $28.8^{+0.9}_{-1.0}$ \\
	Region 2 & $4.29\pm0.04$ & $6.1\pm0.2$ & $6.6\pm0.1$ & $32.2\pm0.9$ \\
	Region 3 & $3.78\pm0.06$& $5.1\pm0.2$ & $3.4\pm0.1$ & $33.4^{+1.3}_{-1.2}$ \\
	Region 4 & $4.20^{+0.07}_{-0.06}$ & $5.3\pm0.2$ & $2.6\pm0.1$ & $28.9^{+1.3}_{-1.1}$ \\
	Region 5 & $3.63\pm0.01$ & $3.4\pm0.2$ & $2.4\pm0.1$ & $23.8\pm1.4$ \\
	Region 6 & $3.51^{+0.09}_{-0.08}$ & $4.1\pm0.3$ & $2.9\pm0.1$ & $30.4\pm1.8$ \\
	Region 8 & $3.76\pm0.06$ & $5.2\pm0.2$ & $3.1\pm0.1$ & $34.2^{+1.4}_{-1.3}$ \\
	Region 9 & $3.19\pm0.06$ & $4.2\pm0.2$ & $2.4\pm0.1$ & $36.7^{+1.8}_{-1.7}$ \\
	Region 10 & $3.56\pm0.05$ & $6.1\pm0.3$ & $5.0\pm0.1$ & $43.7\pm1.8$ \\
	Region 11 & $3.25\pm0.04$ & $4.9\pm0.2$ & $7.2\pm0.1$ & $41.6^{+1.3}_{-1.2}$ \\
	Region 12 & $3.02\pm0.04$ & $4.9\pm0.2$ & $4.5\pm0.1$ & $47.0\pm2.2$ \\ 
	Region 13 & $3.10\pm0.06$ & $3.8^{+0.3}_{-0.2}$ & $5.0\pm0.1$ & $34.8^{+2.6}_{-1.8}$ \\
	Region 14 & $3.34\pm0.06$ & $3.6^\pm0.2$ & $5.0\pm0.1$ & $29.1^{+1.4}_{-1.5}$ \\
	Region 15 & $3.37\pm0.05$ & $4.0\pm0.1$ & $4.1\pm0.1$ & $31.7\pm0.9$ \\
	Region 16 & $3.53\pm0.04$ & $5.4\pm0.2$ & $5.4\pm0.1$ & $39.7^{+1.4}_{-1.3}$ \\
	Region 17 & $3.29\pm0.04$ & $5.3\pm0.1$ & $5.3\pm0.1$ & $43.2\pm1.7$ \\
	Region 18 & $3.22^{+0.06}_{-0.05}$ & $5.4\pm0.3$ & $3.6\pm0.1$ & $45.8^{+2.5}_{-2.4}$ \\
	Region 19 & $3.32\pm0.05$ & $4.9\pm0.2$ & $4.7\pm0.1$ & $39.6\pm1.4$ \\
	Region 20 & $2.81\pm0.07$ & $4.7\pm0.3$ & $1.8\pm0.1$ & $50.1\pm3.6$ \\
	Region 21 & $2.92^{+0.07}_{-0.06}$ & $4.4^{+0.3}_{-0.2}$ & $2.0\pm0.1$ & $44.5^{+3.0}_{-2.3}$ \\
	Region 22 & $3.04\pm0.06$ & $4.5\pm0.2$ & $2.6\pm0.1$ & $42.1^{+1.9}_{-1.8}$ \\
	& & \textbf{Total mass:} & $86.0\pm1.0$ \\
	\hline
	\end{tabular}
	\caption{Density, pressure, mass and entropy index of the X-ray emission in the various regions within the rib structure of Cygnus A. Errors are quoted for 90\% confidence range. These were calculated assuming that ribs are contained to fill a cylinder of radius 39 arcsec.}
	\label{table:sdphysparalt}
\end{table*}

\begin{figure}
	\centering
	\includegraphics[width=0.99\columnwidth]{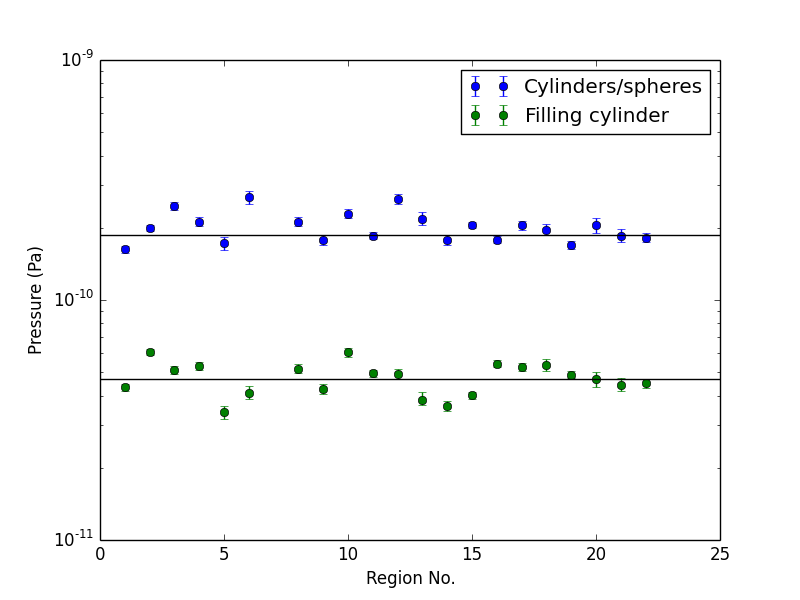}
	\caption{Pressure in each region calculated for the two different geometries. Lines of best fit indicate the mean for each pressure group.}
	\label{fig:pressurecomp}
\end{figure}

A comparison between the pressures for each volume assumption can be seen in Figure \ref{fig:pressurecomp}. Fitting a constant to these data gives an average pressure $\overline{P} = (2.0\pm0.1) \times 10^{-10}$ Pa with $\chi^{2}$ = 246.4 for the cylinders and spheres and $\overline{P} = (4.7\pm0.2) \times 10^{-11}$ Pa with $\chi^{2}$ = 234.7 assuming the ribs fill a cylinder. This fitting shows no significant preference for either model. The clear difference in pressure between the two volume models suggests that the ribs are not composed of a mixture of cylinders and spheres and regions which fill a cylinder. This suggests that all structures within the ribs are likely to have a similar geometry and similar dynamical origins. The poor $\chi^{2}$ and deviations from the mean are possibly caused by inhomogeneities in the density of the various rib structures. For both geometries the total mass of rib gas is substantial at a few times $10^{10} M_{\odot}$.

\mbox{\citet{2017Snios}} have determined the pressure of the lobes to be $(1.0\pm0.4)\times10^{-10}$ Pa in the east lobe and $(8.4\pm0.2)\times10^{-11}$ Pa in the west lobe using spectra for the thin rim of gas between the shock and the radio lobes. This is an order of magnitude larger than the $1.3\times10^{-11}$ Pa minimum energy field of lobes closest to the centre of the source found by \citet{1991Carilli}. The mean pressures we find in our two configurations fall either side of the Snios et al. estimates.

\subsection{Dynamical timescales of the ribs}
\label{sec:dyntime}
The sound speed, $c_{s}$, of the gas in each region is found using Equation \ref{eqn:cs} \citep{2012Worrall}, where $T$ is the temperature of the regions obtained from our spectral fits.
\begin{align}
\label{eqn:cs}
\centering
c_{s} (\rm{kpc\,Myr}^{-1}) = 0.54 (\emph{kT}/\rm{keV})^{0.5}
\end{align}
We can estimate the dynamical timescale of each region by taking the length of the shortest two-dimensional axis and dividing it by the calculated sound speed. If the source is older than this timescale the gas in each region is likely to be in pressure balance. The results for $c_{s}$ and the dynamical timescale are given in Table \ref{table:csage}. The timescales vary between 3.1 and 9.7 Myr, and are somewhat less than the 18.7 Myr estimated age of the shock front \citep{2017Snios} implying that the structure has largely reached local pressure equilibrium.

\begin{table}
	\begin{tabular}{l c c}
	\hline
	Region & $c_{s}$ & Dynamical   \\
	& (kpc Myr$^{-1}$) & timescale (Myr) \\
	\hline
	Region 1 & $0.97\pm0.02$ & $6.3\pm0.1$ \\
	Region 2 & $1.07\pm0.02$ & $9.7\pm0.1$ \\
	Region 3 & $1.05\pm0.02$ & $3.9\pm0.1$ \\
	Region 4 & $1.01\pm0.01$ & $6.6\pm0.1$ \\
	Region 5 & $0.87\pm0.02$ & $4.7\pm0.2$ \\
	Region 6 & $0.98\pm0.03$ & $3.1\pm0.1$ \\
	Region 8 & $1.06\pm0.02$ & $6.8\pm0.1$ \\
	Region 9 & $1.04\pm0.02$ & $6.7\pm0.1$ \\
	Region 10 & $1.17\pm0.02$ & $4.8\pm0.1$ \\
	Region 11 & $1.11\pm0.01$ & $5.7\pm0.1$ \\
	Region 12 & $1.15\pm0.03$ & $3.4\pm0.1$ \\ 
	Region 13 & $1.00^{+0.04}_{-0.02}$ & $3.7\pm0.1$ \\
	Region 14 & $0.94\pm0.02$ & $5.3\pm0.1$ \\
	Region 15 & $0.98\pm0.01$ & $5.3\pm0.1$ \\
	Region 16 & $1.12^{+0.02}_{-0.01}$ & $7.6\pm0.1$ \\
	Region 17 & $1.14\pm0.01$ & $6.3\pm0.1$ \\
	Region 18 & $1.16\pm0.03$ & $4.9\pm0.1$ \\
	Region 19 & $1.09\pm0.02$ & $8.9\pm0.1$ \\
	Region 20 & $1.16\pm0.04$ & $5.5\pm0.2$ \\
	Region 21 & $1.11^{+0.04}_{-0.03}$ & $6.1^{+0.2}_{-0.1}$ \\
	Region 22 & $1.09\pm0.02$ & $6.8\pm0.1$ \\
	\hline
	\end{tabular}
	\caption{Sound speed and age of the X-ray gas contained within each region of the rib structure in Cygnus A.}
	\label{table:csage}
\end{table}

\section{Mapping}
\subsection{Temperature Mapping}
\label{sec:tempmap}
\begin{figure}
	\centering
	\includegraphics[width=0.99\columnwidth]{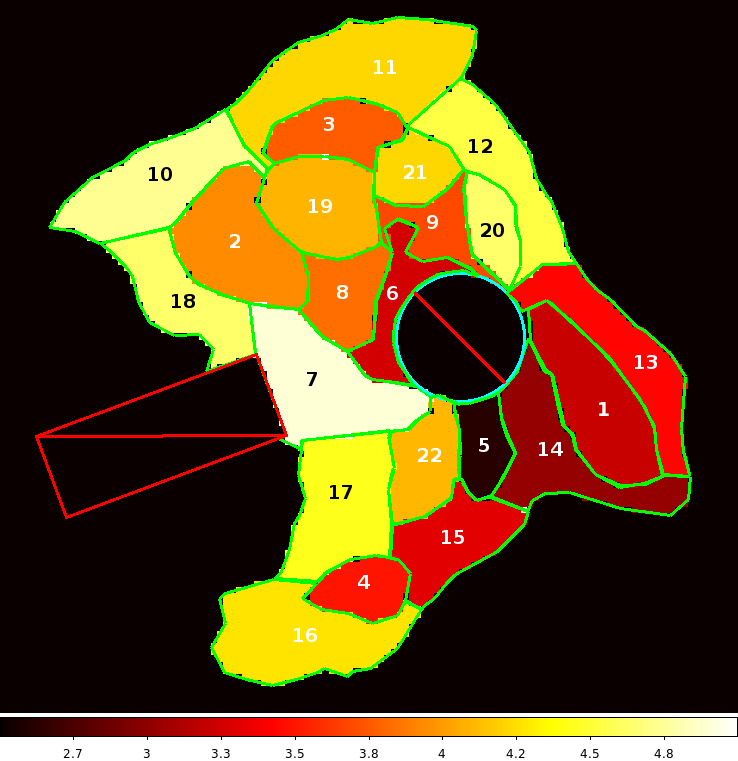}
	\caption{Temperature map of the rib-like structure of Cygnus A. The colourbar corresponds to the best-fit temperature of each region in keV. The black circle at the centre of the sources corresponds to an excised circle of radius 5 arcsec, representing the core.}
	\label{fig:sdtemp}
\end{figure}

The best-fit temperature map in the rib gas is shown in Figure \ref{fig:sdtemp}. 
Interestingly, there is a clear range in temperature within the structure, with regions in the southwestern corner of the ribs cooler than those in the remainder, although the temperatures in regions 6 and 13 have been reduced by 0.5 and 0.4 keV respectively (more than 10 per cent) by the inclusion of the core component.

\begin{figure}
	\centering
	\includegraphics[width=0.99\columnwidth]{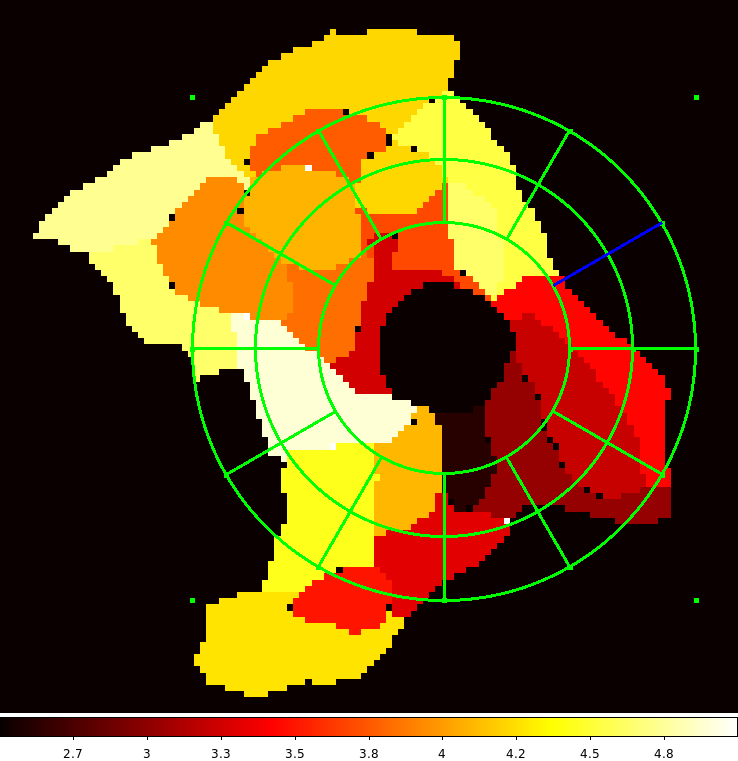}
	\caption{Temperature map of the ribs showing the pie slices used to model the temperature at various angles and radii. The blue line indicates the 0 degree line.}
	\label{fig:pandas}
\end{figure}

\begin{figure}
	\centering
	\includegraphics[width=0.99\columnwidth]{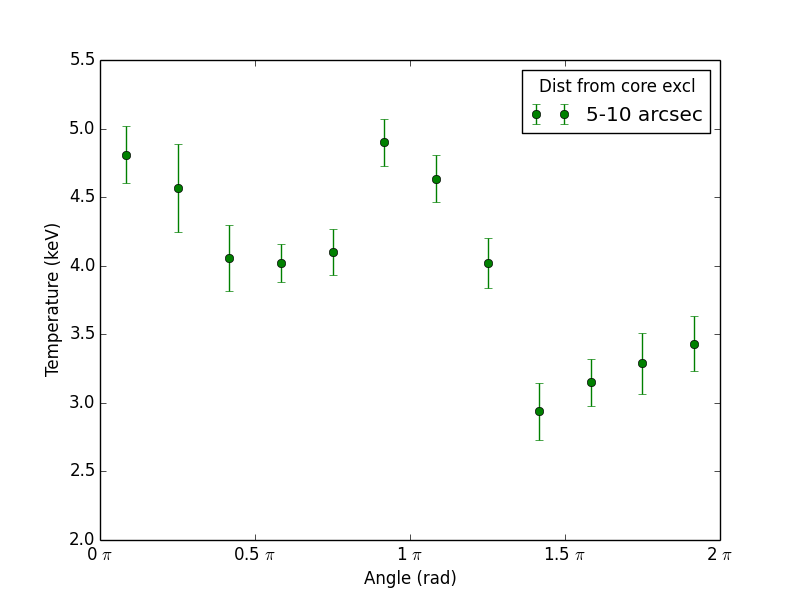}
	\includegraphics[width=0.99\columnwidth]{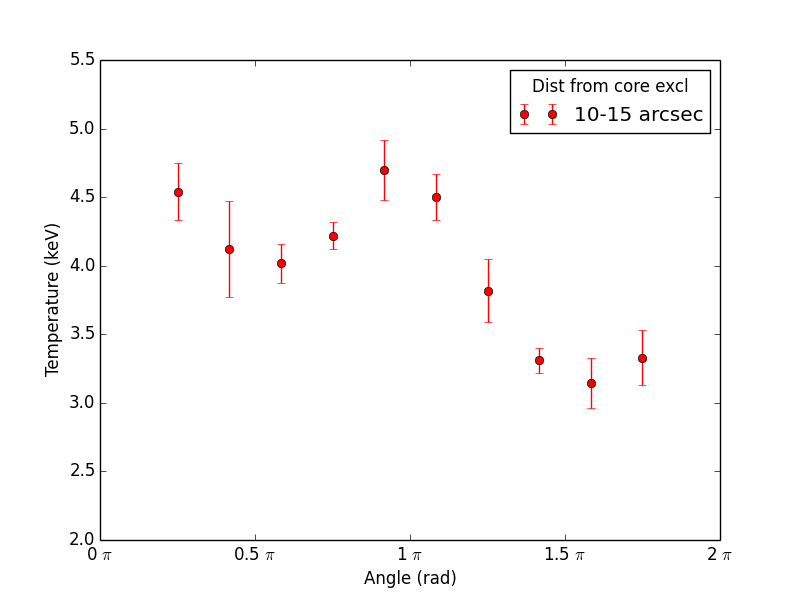}
	\caption{Temperature versus angle across the rib-like structure in Cygnus A between 5-10 arcseconds from the core exclusion (middle) and 10-15 arcseconds from the core exclusion (bottom). The first and final wedges of the 10-15 arcsecond plot are missing as there is no rib structure in these segments at this radius.}
	\label{fig:dipole}
\end{figure}

To obtain a measure of the range in temperature, we placed annuli of width 5 arcseconds centred on the core and calculated the average temperature in 30 degree annular sectors (see Figure \ref{fig:pandas}). The results for annuli at 5-10 arcseconds and 10-15 arcseconds can be seen in Figure \ref{fig:dipole}. The difference between the hottest and coolest regions is about 2 keV. The temperature difference is compensated by an opposing density asymmetry that causes the pressure to be roughly constant in the ribs. 

\subsection{Pressure and Entropy Index Maps}
\label{sec:pandsmaps}

\begin{figure}
	\centering
	\includegraphics[width=0.99\columnwidth]{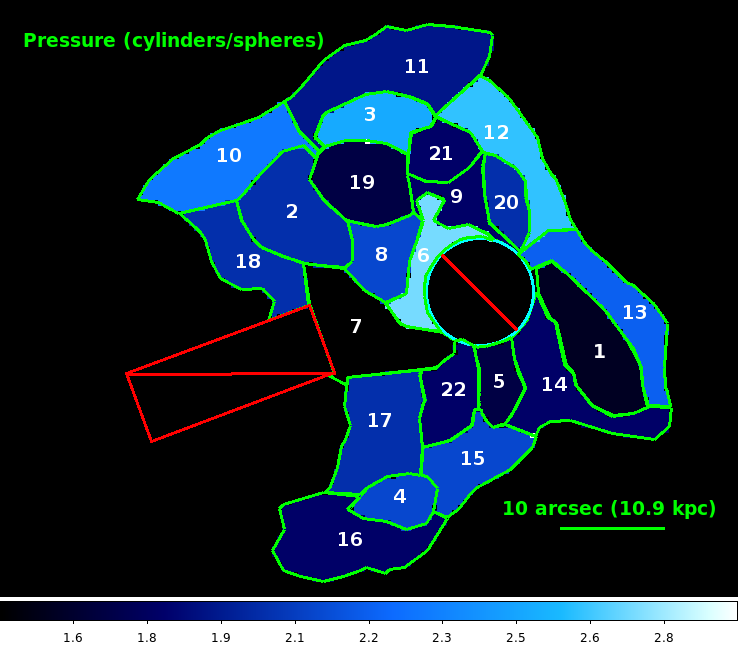}
	\includegraphics[width=0.99\columnwidth]{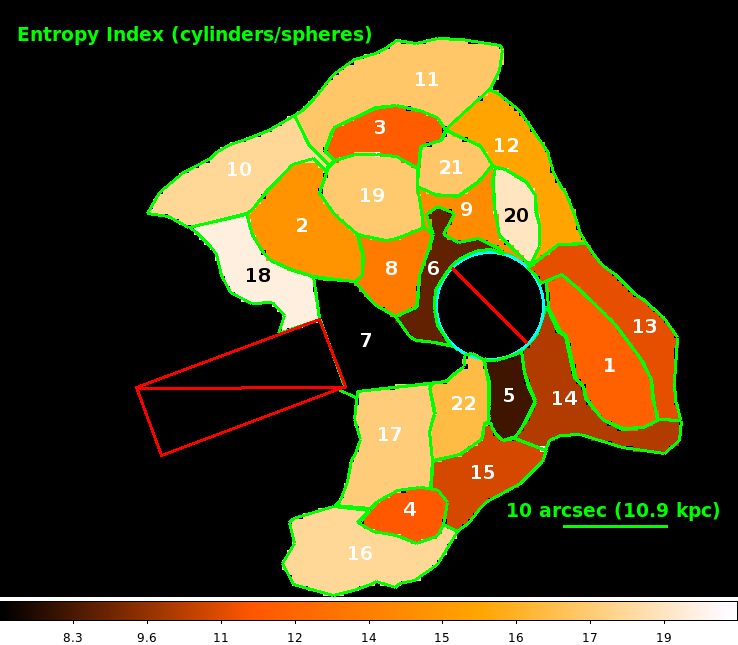}
	\caption{\emph{Top:} Pressure map of the rib-like structure of Cygnus A, where volumes are calculated using cylinders and spheres representing the regions. The colourbar corresponds to the best-fit pressure of each region in $10^{-10}$ Pa. The black circle at the centre of the sources corresponds to an excision of radius 5 arcsec, representing the core. \emph{Bottom:} Entropy index map of the rib-like structure of Cygnus A, where volumes are calculated using cylinders and spheres representing the regions. The colourbar corresponds to the entropy index of each region in keV cm$^{2}$. The black circle at the centre of the sources corresponds to an excision of radius 5 arcsec, representing the core.}
	\label{fig:presmap}
\end{figure}

\begin{figure}
	\centering
	\includegraphics[width=0.99\columnwidth]{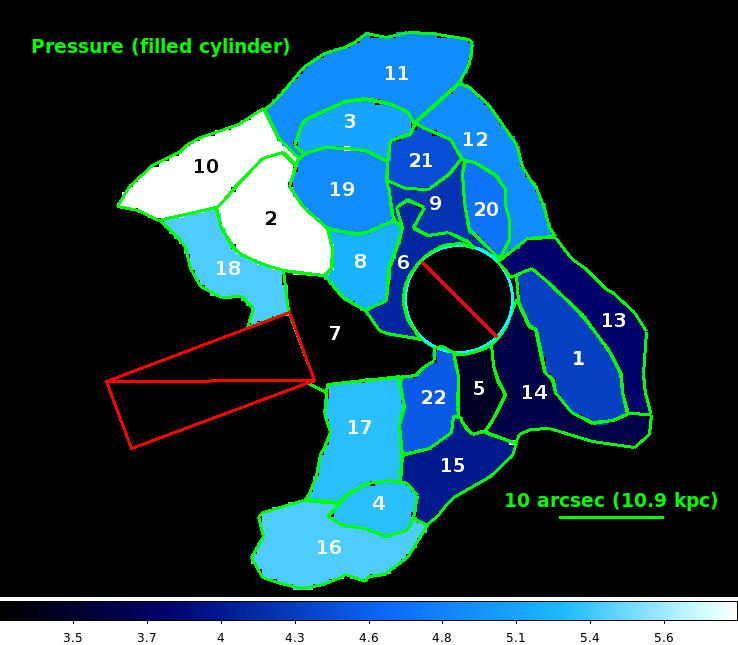}
	\includegraphics[width=0.99\columnwidth]{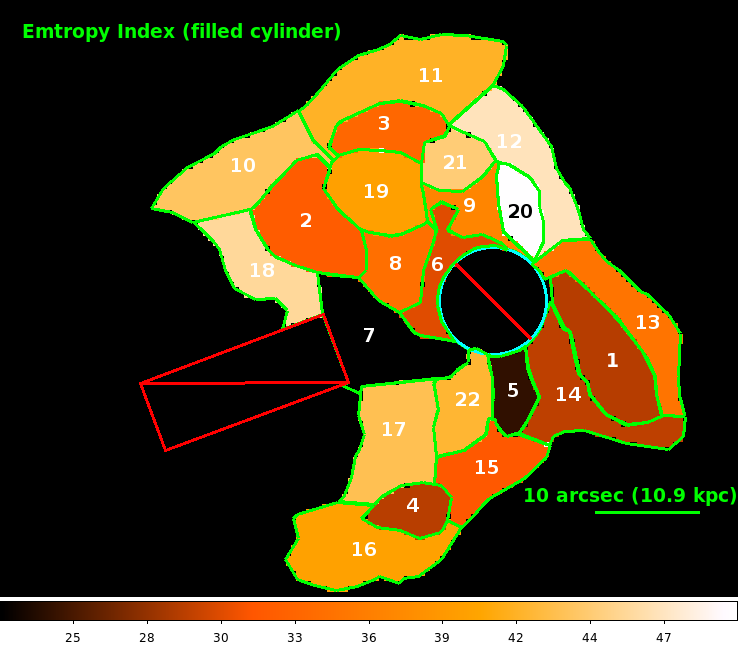}
	\caption{\emph{Top:} Pressure map of the rib-like structure of Cygnus A, where volumes are calculated assuming the ribs fill a cylinder defined by the cocoon shock. The colourbar corresponds to the best-fit pressure of each region in $10^{-11}$ Pa. The black circle at the centre of the sources corresponds to an excision of radius 5 arcsec, representing the core. \emph{Bottom:} Entropy index map of the rib-like structure of Cygnus A, where volumes are calculated assuming the ribs fill a cylinder defined by the cocoon shock. The colourbar corresponds to the entropy index of each region in keV cm$^{2}$. The black circle at the centre of the sources corresponds to an excision of radius 5 arcsec, representing the core.}
	\label{fig:prescylmap}
\end{figure}

Assuming that the ribs are composed of cylinders and spheres, there is no discernible systematic structure in the pressure map (see Figure \ref{fig:presmap}). However, in Figure \ref{fig:prescylmap} (ribs filling cylinder) the regions in the eastern half of the ribs contain somewhat higher pressure when compared to the western half, which is a pattern also identified for the lobes \citep{2017Snios}. The average pressures of the eastern half of the ribs, for this geometry, is $(5.5\pm0.2)\times10^{-11}$ Pa and the average pressure of the western half is $(4.4\pm0.2)\times10^{-11}$ Pa, which is 10 per cent lower. While less significant than the roughly 20 per cent difference seen by Snios et al., it is interesting that it an east-west pressure difference may be present closer to the nucleus. However, pressures in the east and west of the cylinders and spheres configuration are the same within errors.

We also investigate the entropy index of the ribs, which is of interest as the entropy structure contains information about the thermal history of the gas in clusters, especially with regard to the effect of feedback on the intracluster medium (ICM). In general, the entropies across the ribs are uniform, with the highest entropy index for both geometries (lower images, Figures \ref{fig:presmap} and \ref{fig:prescylmap}) found at the extremities of the rib structure. The exception to this is the southwestern corner and entropies here are similar to those seen in regions 2 and 3. Interestingly, the entropies in regions 5, 15 and 4, which arguably belong to the same rib, are similar but increase with angle from the core. Perhaps unsurprisingly, the entropies measured for the filled cylinder geometry are closer to the central entropy of Cygnus A found by \citet{2009Cavagnolo}, assuming spherical symmetry.

Although we have discussed differences in pressure and entropy in the rib structure, within each geometry the pressures and entropies measured are similar across regions. There is less than a factor of 2 difference between the lowest and highest pressures and entropies within each geometry. As it is unlikely the two geometries are mixed (see Figure \ref{fig:pressurecomp}), and because of our suggested origin for the ribs (see Section \ref{sec:origin}) we prefer the cylinder and spheres geometry for the following sections.

\section{Rayleigh-Taylor Instabilities}
\label{sec:rt}
The gas extending to the south of the core is one of the more striking
parts of Cygnus~A's ribs. The eastern edge of region~5 (Figure~\ref{fig:regselfdefined}) is a
particularly sharp boundary. This is remarkable since it appears that
X-ray emitting gas is decelerating the low-density plasma from a part
of the Cygnus~A lobe system, and so this interface should be subject
to the Rayleigh-Taylor (RT) instability. In this Section we discuss
this edge of region~5 as an exemplar of other sharp features in
Cygnus~A (Figure~\ref{fig:unsharpmask}). 

RT instabilities develop where a higher-density gas lies above a
lower-density gas in a gravitational field, or, equivalently, where
the higher-density gas is decelerating the expansion of the
lower-density gas. Bubbles and fingers of the fluids then disturb the
interface and cause mixing. If the lower-density gas lies over the
higher-density gas, or if the higher-density gas is accelerating the lower-density gas, then there
are no RT instabilities. In the context of Cygnus~A, we expect RT
instabilities where radio plasma is encountering denser external
gas.

In the absence of dynamically-important magnetic fields 
the growth rate of the RT instability is (Chandrasekhar 1961)
\begin{align}
  \Gamma = \left( 2\pi\frac{g}{\lambda}
           \frac{\rho_{\rm a} - \rho_{\rm b}}{\rho_{\rm a} + \rho_{\rm b}}
           \right)^{1/2}
  \label{eq:grbasic}
\end{align}
where $g$ is the deceleration, $\rho_a$ and $\rho_b$ are the densities
of the two gases ($\rho_a > \rho_b$), and $\lambda$ is the horizontal
(parallel to the interface) wavelength of a mode. RT instabilities
grow at all wavelengths, but faster at shorter $\lambda$.

The density of the radio-emitting plasma in Cygnus~A is much lower
than the density of the external rib gas, $\rho_{\rm a} \gg \rho_{\rm b}$,
so the growth rate of the instability simplifies to
\begin{align}
  \Gamma = \left(2\pi\frac{g}{\lambda}\right)^{1/2} \quad .
  \label{eq:grapprox}
\end{align}
If the instabilities have been growing for time, $\tau$, then any
initial perturbations in the interface will grow by a factor
$\exp(\Gamma \tau)$
which should be $< 1$ for the perturbations still to be in the
linear regime. If the radio plasma has been interacting with the X-ray
plasma for the same time $\tau$, during which the lobe has expanded to
scale $R$, then $g$ will be of order $R/\tau^2$ (e.g., for
self-similar expansion), and 
\begin{equation}
  \exp(\Gamma \tau) \approx \exp\left( 2.5 \left( \lambda/R
    \right)^{-{1/2}} \right) \quad . 
  \label{eq:growthinviscid}
\end{equation}
Perturbations on scales smaller than the radio structure should
therefore have grown to the non-linear regime and blurred the
interface, unless the radio lobe lies over the thermal plasma, or the
thermal plasma is accelerating the radio lobe.

However, such a conclusion is premature, since the result
(\ref{eq:growthinviscid}) is unrealistic. We expect magnetic
fields in the radio plasma and thermal gas to become concentrated and
ordered near the interface and that the magnetic layer will inhibit
the RT instability (e.g., \citealt{2017Carlyle}). A full
treatment of the RT instability then requires a model for the magnetic structure. Since the magnetic field
is expected to lie predominantly parallel to the interface, but with
no consistent direction on the large scale, and with the field
strength decreasing away from the interface, no analytic or numerical
calculation to date provides a sufficient treatment of the growth of
the instability. However, an approximate result (validated by a
comparison with simpler field models, such as that of \citealt{1995Goldston}) 
can be obtained by modelling the field as providing
surface energy that mimics surface tension. Surface tension, 
$\sigma_T$, introduces a wavelength of maximum instability 
\begin{equation}
  \lambda_{\ast} = 2 \pi \left( \frac{3 \sigma_T}{( \rho_{a} - \rho_{b} ) g}
    \right)^{1/2} 
  \label{eq:lambdastar}
\end{equation}
\noindent 
and suppresses the RT instability completely for modes with
$\lambda < \lambda_\ast/\sqrt{3}$.
The rate of growth of the RT instability is modified to
\begin{equation}
  \Gamma = \left[ 2 \pi \frac{g}{\lambda} \Big\{ \frac{\rho_a -
    \rho_b}{\rho_a + \rho_b} - \frac{ 4 \pi^2 \sigma_T}{g \lambda^2
      (\rho_a + \rho_b)} \Big\} \right]^{1/2}
  \label{eq:gammawithst}
\end{equation}
and takes a maximum value of
\begin{equation}
  \Gamma_\ast = \left[ \frac{4 g^3 ( \rho_a - \rho_b)^{3}}
    { 27 \sigma_T ( \rho_a + \rho_b )^2 } \right]^{1/4} 
  \label{eq:gammastar}
\end{equation}
at $\lambda = \lambda_\ast$.

The model quantity $\sigma_T$ is taken as arising from
magnetic field energy density $u_B$ concentrated in a layer of
thickness $d$ at the edge of the radio lobe. $d$ could be the vertical 
scale on which the RT instability causes field bunching, or some
scale established by the dynamics of the expansion of the radio
lobe. The effective surface tension is then
\begin{align}
  \sigma_T = u_{B} d \quad .
  \label{eq:surften}
\end{align}
We take the value of $u_B$ as some multiple, $f_B$, of the
minimum-energy pressure in the lobes of Cygnus~A, so that 
$u_B = 3 f_B \times 10^{-10} \ \rm J \, m^{-3}$. 
The gas in region~5 has
$\rho_{a} \approx 3 \times 10^{-22} \ \rm kg \, m^{-3}$ (Table~3). 
Equation (\ref{eq:lambdastar}) then gives a scale of maximum instability
\begin{equation}
  \lambda_\ast \approx 40 f_B^{1/2} \left( d / {\rm kpc} \right)^{1/2}
    \ \rm kpc
  \label{eq:maglambda}
\end{equation}
which exceeds the radius $R\sim30$ kpc of Cygnus A's cocoon shock
unless the magnetic layer is thin or $f_{B}$ is small. That is, a magnetic 
layer of thickness > 0.1 kpc at the edge of the radio plasma bubble east of 
region~5, can fully stabilise the interface against RT instabilities with local magnetic field
enhancements of less than a factor of 3 ($f_{B} \lesssim 10$). The field is then less than 90 nT.

\begin{figure}
  \centering
  \includegraphics[width=0.99\columnwidth]{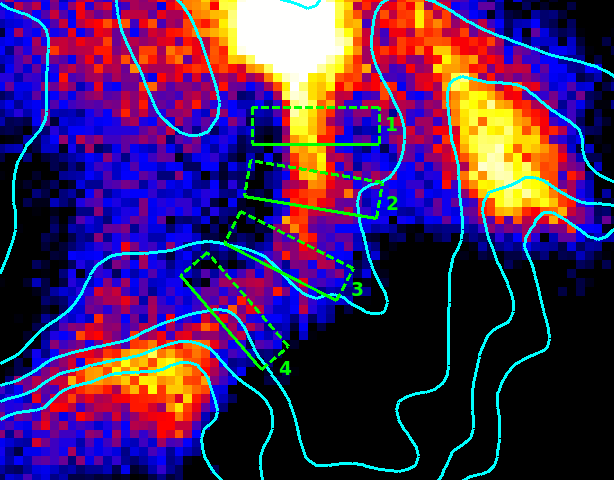}
  \caption{Merged \emph{Chandra} event file image of Cygnus A
     observations listed in Table \ref{table:obs}. with native
     0.492 arcsec pixels, zoomed in on the south rib and
     associated arm with radio contours from Figure \ref{fig:xrayobs} 
     overlaid in cyan. Each box indicates the position of extracted
     counts profiles. Boxes are of width 2 arcseconds with counts
     in the profiles averaged over the thickness of the box.} 
  \label{fig:proj_regions}
\end{figure}

We quantified the sharpness of the interface at region~5 by measuring
the width of count profiles perpendicular to the interface at four
positions along its length (see Fig.~\ref{fig:proj_regions}).
We specified the interface width as the distance
between counts corresponding to 0.25 and 0.75 of the maximum
above the surrounding background. We call this measured width
$\sigma_{\rm obs}$, where 
$\sigma_{\rm obs}^{2} = \sigma_{\rm intrinsic}^{2} + \sigma_{\rm PSF}^{2}$. 
We found $\sigma_{\rm PSF}$ based a simulated PSF
generated using the \emph{Chandra} 
software \texttt{ChaRT} and \texttt{MARX}. We defined the spectrum
using the data for region~5, and projected the ray tracings
onto the ACIS-I detector at the centre of region~5 on ObsID 5831. Our
PSF is a sum of 100 simulated PSFs giving a total exposure of over 5
Ms. $\sigma_{\rm PSF}$, the width between 0.25 and 0.75 of the maximum
counts of the PSF, is measured to be 0.43 arcsec (470 kpc). 

\begin{figure*}
  \begin{tabular}{cc}
    \subfloat[Profile 1; After PSF correction, $\sigma_{\rm intrinsic}=460$
      pc (0.42 arcsec)]{\includegraphics[width=0.5\textwidth]{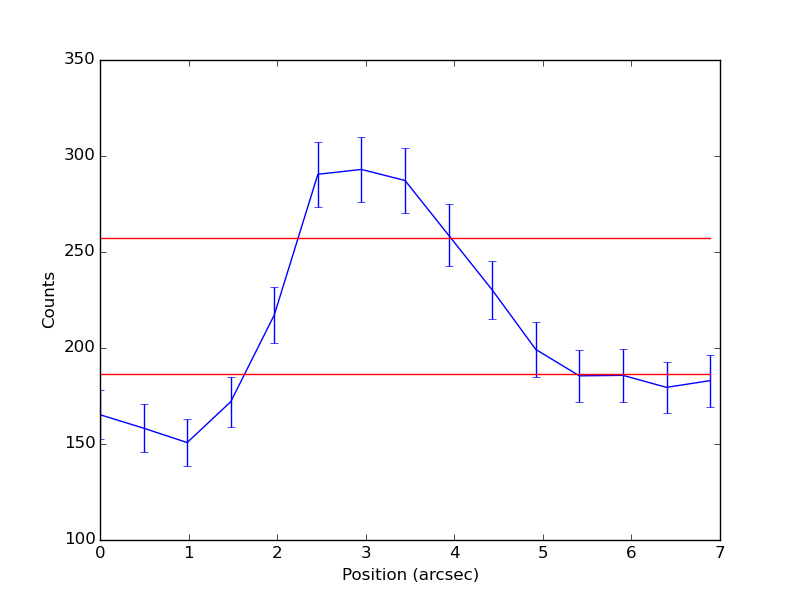}} & 
    \subfloat[Profile 2; After PSF correction, $\sigma_{\rm intrinsic}=880$
      pc (0.81 arcsec)]{\includegraphics[width=0.5\textwidth]{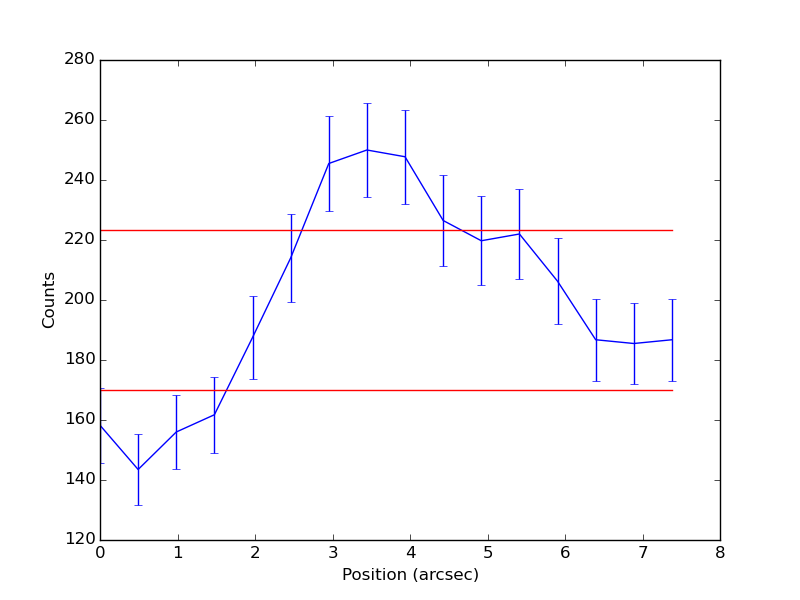}} \\ 
    \subfloat[Profile 3; After PSF correction $\sigma_{\rm intrinsic}=400$
      pc (0.37 arcsec)]{\includegraphics[width=0.5\textwidth]{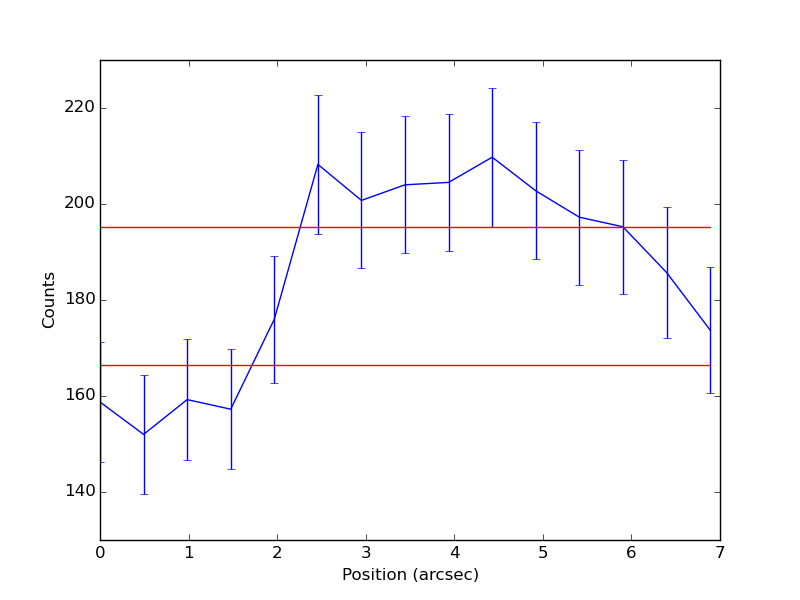}} & 
    \subfloat[Profile 4; After PSF correction, $\sigma_{\rm intrinsic}=370$
      pc (0.34 arcsec)]{\includegraphics[width=0.5\textwidth]{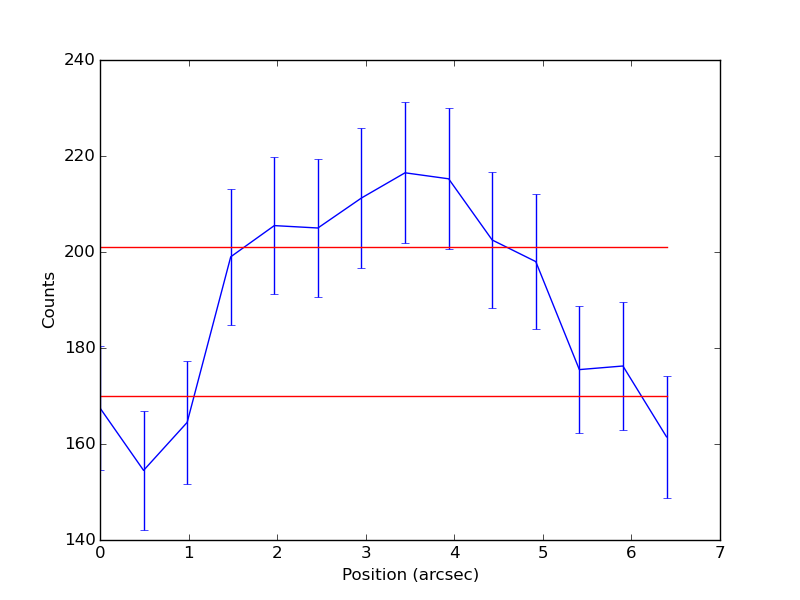}} \\ 
  \end{tabular}
  \caption{Count profiles of the various regions along the south
    rib. Red lines indicate the 0.25 and 0.75 of the maximum counts
    above the background.} 
  \label{fig:profiles}
\end{figure*}

The count profiles from each of the projection boxes are shown in
Figure \ref{fig:profiles}, along with the estimated upper limit on
$\sigma_{\rm intrinsic}$, and show a typical PSF-corrected 
interface width of around 400~pc.

For a mode of $\lambda \approx 400$ pc to be unstable, $d < 0.2
f_B^{-1}$~pc. Such a small scale for a magnetic layer seems
implausible. Thus we expect that the RT instability at the edge of
region~5, and by extension at the other sharp edges in Cygnus~A, will
be suppressed by the effective surface tension of magnetised layers at
the edge of the radio lobe, even if the plasmas are in an RT-unstable
configuration. 

It should be noted that adding a surface-tension-like energy in
magnetic field to the edges of the radio lobes in this way has little
effect on the estimated lobe pressure since the pressure differential
introduced is a factor 
$2 f_B d / R$ less than the minimum energy pressure, and this factor
is small for any plausible magnetic layer.

\section{An Origin for the Ribs} 
\label{sec:origin}
It is unlikely that the ribs are formed from a large-scale accretion event with gas flowing into the Cygnus A nucleus as suggested by \citet{2002Smith}. The amount of mass we calculate is of order $10^{10} M_{\odot}$, which is much larger than expected in accretion disks. \citet{2010Mathews} suggest that the ribs may be the expanded remnant of denser gas formerly located near the centre of Cygnus A, which has been shocked and heated by the AGN during the early stages of FRII development. The uniform low entropies seen across the rib structure imply that not a lot of strong shock heating has occurred within the structure.

Because of the total mass, morphology and dynamical ages of the ribs, we suggest that they result from the destruction of the cool core of the Cygnus A cluster during the early stages of jet propagation. After the radio outburst was triggered, the radio jets would have passed through the cool core, heating the gas and perhaps causing it to fragment and expand if the break out is slow. Backflow antiparallel to the jet's motion after the jets exit the cool core could cause the fragments to roll up into the cylindrical filaments, as is seen in simulations with cylindrical symmetry \citep{2010Mathews,2012Mathews}. The ribs are then lifted to the central inside surface of the cocoon rim by the pressure of the underlying structure. 

It is possible that a minor merger between the Cygnus A host galaxy and another galaxy had some impact in shaping the ribs. The favoured explanation for a transient detected in both K-band observations from Keck II \citep{2003Canalizo} and new JVLA observations \citep{2017Perley} is that it represents a secondary supermassive black hole in orbit around the Cygnus A primary. It is speculated that this merger delivered the gas which caused the ignition of the radio source in Cygnus A, and is also the origin of the transient. However, while it is possible that a merger had some impact on the shaping of ribs, it is not the overall origin of the gas, the large mass we measure argues against the ribs originating as gas stripped from the interstellar medium (ISM).

Our interpretation suggests that the volume of the ribs is more likely to be represented by the projected cylinders and spheres, than a geometry that fills the cocoon. This gives a total mass in the structure of $(2.1\pm0.1)\times10^{10} M_{\odot}$, which matches well with the gas masses found in the cool cores of other clusters \citep{2006Vikhlinin,2017Hogan}. The rib structure is a similar size to the cool cores of other clusters, measuring around 50 kpc across along its north-south axis \citep{2006Vikhlinin,2007Pratt}. This model suggests that the pressures and entropies across the rib structure should vary by only modest functions (as seen across smoother structures in other cluster cool cores by \citealt{2014Panagoulia}), and those which we measure vary only by a factor of 2. Pressure equilibrium with the lobes means that the ribs are no longer undergoing local expansion and will expand and contract with the cocoon.

The southwestern temperature deficit seen in Figure \ref{fig:sdtemp} could result if the host galaxy of Cygnus A was a few kpc to the northeast of the cluster centre at the start of the radio outburst. BCGs are frequently found to be off-centre as a result of mergers \citep{2014Martel} and Cygnus A is thought to be undergoing a merger currently with a sub-cluster to its northwest \citep{2017Wise}.

Parts of this explanation and the lack of gas mass transport to outer regions seen here are similar to results of simulations of GHz-peak spectrum objects undertaken by \citet{2007Sutherland}. Obviously, those sources are far smaller than Cygnus A (around 60 times) and have disrupted their ISMs rather than a cluster cool core as we infer for Cygnus A.

The rib filaments may be associated with enhanced magnetic fields. This, in combination with the already clear density variations in the region, suggests there would be rotation measure structures present across the central region. Unfortunately, the rotation measure maps of \citet{1987Dreher}, which find a RMs of up to a few thousand rad m$^{-2}$, do not cover the central regions containing the ribs. There is a significant amount of gas in the region, indicating the rotation measure could be large, perhaps up to 10,000 rad m$^{-2}$ depending on the degree of magnetic ordering in the interfaces. Radio telescopes such as LOFAR and JVLA could provide measurements of such high RMs and RM gradients near the ribs, but these measurements will be challenging: they must be done at low frequencies so that synchrotron emission in these regions is bright; at high angular resolution to resolve the cavity edges; at high spectral resolution to avoid depolarisation; and at high dynamic range. The question of whether the thermal gas associated with the ribs is in the interior regions of the radio lobes should also be answered by radio polarimetry.

\section{Summary}
We have used X-ray spectroscopy from very deep \emph{Chandra} observations to investigate the rib-like structure within the cocoon shock of Cygnus A and have confirmed it to be dominated by thermal emission with an average temperature of about 4 keV (Section \ref{sec:spectralfitting}, Table \ref{table:sdspecfits}).

Dynamical timescales for the ribs are somewhat less than the estimated age of the shock front, suggesting that the gas is in local pressure equilibrium with its surroundings and expanding with the cocoon (Section \ref{sec:dyntime}, Table \ref{table:csage}). Similarities in the pressures across the rib structure suggest that it is a coherent structure with a single dynamical origin (Sections \ref{sec:physpar} and \ref{sec:pandsmaps}).

The clumpy, filamentary, rib-like structure is likely debris resulting from disintegration of the cool core of the Cygnus A cluster during the early stages of the current epoch of activity (Section \ref{sec:origin}). The gas now lies on the central inside surface of the cocoon rim (Section \ref{sec:origin}). The pressure across the ribs is roughly constant if the filaments are modelled as structures lying within a sheath; this is not the case if some are elongated structures projected to fill the radio lobes.

We have discovered that the gas exhibits a temperature structure, with the southwestern part of the ribs cooler than the rest (Figure \ref{fig:sdtemp}). Plotting the temperature as a function of angle 5-10 arcseconds and 10-15 arcseconds from the core shows an oscillatory pattern with an amplitude of about 1 keV providing a quantitative measure of the temperature gradient (Figure \ref{fig:dipole}). We suggest that the X-ray gas in the southwest region is cooler and the AGN was a few kpc northeast of the cluster centre at the start of the outburst.

RT instabilities are not active in the sharp edge measured in detail for one of the gas filaments. We show that they could have been suppressed by magnetised layers at the edge of the radio lobe (Section \ref{sec:rt}).

\section*{Acknowledgements}
Support for this work was provided by the National Aeronautics and Space Administration through Chandra
Award Number GO5-16117A issued by the Chandra X-ray Observatory Center, which is operated by the Smithsonian Astrophysical Observatory for and on behalf of the National Aeronautics Space Administration
under contract NAS8-03060. RTD thanks STFC for support. PEJN was supported in part by NASA contract NAS8-03060. MJH acknowledges STFC grant ST/M001008/1. ACE acknowledges support from STFC grant ST/L00075X/1.

\bibliography{bib}

\newcommand{\noop}[1]{}
\begin{thebibliography}{}

\bibitem[\protect\citeauthoryear{{Anders} \& {Grevesse}}{{Anders} \&
  {Grevesse}}{1989}]{1989Anders}
{Anders} E.,  {Grevesse} N.,  1989, \gca, 53, 197

\bibitem[\protect\citeauthoryear{{Arnaud}}{{Arnaud}}{1996}]{1996Arnaud}
{Arnaud} K.~A.,  1996, in {Jacoby} G.~H.,  {Barnes} J.,  eds, Astronomical Data
  Analysis Software and Systems V Vol.~101 of Astronomical Society of the
  Pacific Conference Series, {XSPEC: The First Ten Years}.
p.~17

\bibitem[\protect\citeauthoryear{{Arnaud}, {Fabian}, {Eales}, {Jones} \&
  {Forman}}{{Arnaud} et~al.}{1984}]{1984Arnaud}
{Arnaud} K.~A.,  {Fabian} A.~C.,  {Eales} S.~A.,  {Jones} C.,    {Forman} W.,
  1984, \mnras, 211, 981

\bibitem[\protect\citeauthoryear{{Canalizo}, {Max}, {Whysong}, {Antonucci} \&
  {Dahm}}{{Canalizo} et~al.}{2003}]{2003Canalizo}
{Canalizo} G.,  {Max} C.,  {Whysong} D.,  {Antonucci} R.,    {Dahm} S.~E.,
  2003, \apj, 597, 823

\bibitem[\protect\citeauthoryear{{Carilli}, {Perley}, {Dreher} \&
  {Leahy}}{{Carilli} et~al.}{1991}]{1991Carilli}
{Carilli} C.~L.,  {Perley} R.~A.,  {Dreher} J.~W.,    {Leahy} J.~P.,  1991,
  \apj, 383, 554

\bibitem[\protect\citeauthoryear{{Carlyle} \& {Hillier}}{{Carlyle} \&
  {Hillier}}{2017}]{2017Carlyle}
{Carlyle} J.,  {Hillier} A.,  2017, \aap, 605, A101

\bibitem[\protect\citeauthoryear{{Cavagnolo}, {Donahue}, {Voit} \&
  {Sun}}{{Cavagnolo} et~al.}{2009}]{2009Cavagnolo}
{Cavagnolo} K.~W.,  {Donahue} M.,  {Voit} G.~M.,    {Sun} M.,  2009, \apjs,
  182, 12

\bibitem[\protect\citeauthoryear{{Chon}, {B{\"o}hringer}, {Krause} \&
  {Tr{\"u}mper}}{{Chon} et~al.}{2012}]{2012Chon}
{Chon} G.,  {B{\"o}hringer} H.,  {Krause} M.,    {Tr{\"u}mper} J.,  2012, \aap,
  545, L3

\bibitem[\protect\citeauthoryear{{de Vries} et~al.,}{{de Vries}
  et~al.}{submitted}]{2017deVries}
{de Vries} M.~N.,  et~al., submitted, \mnras

\bibitem[\protect\citeauthoryear{{Dickey} \& {Lockman}}{{Dickey} \&
  {Lockman}}{1990}]{1990Dickey}
{Dickey} J.~M.,  {Lockman} F.~J.,  1990, \araa, 28, 215

\bibitem[\protect\citeauthoryear{{Dreher}, {Carilli} \& {Perley}}{{Dreher}
  et~al.}{1987}]{1987Dreher}
{Dreher} J.~W.,  {Carilli} C.~L.,    {Perley} R.~A.,  1987, \apj, 316, 611

\bibitem[\protect\citeauthoryear{{Duffy}, {Worrall}, {Birkinshaw} \&
  {Kraft}}{{Duffy} et~al.}{2016}]{2016Duffy}
{Duffy} R.~T.,  {Worrall} D.~M.,  {Birkinshaw} M.,    {Kraft} R.~P.,  2016,
  \mnras, 459, 4508

\bibitem[\protect\citeauthoryear{{Fabian}, {Sanders}, {Ettori}, {Taylor},
  {Allen}, {Crawford}, {Iwasawa} \& {Johnstone}}{{Fabian}
  et~al.}{2001}]{2001Fabian}
{Fabian} A.~C.,  {Sanders} J.~S.,  {Ettori} S.,  {Taylor} G.~B.,  {Allen}
  S.~W.,  {Crawford} C.~S.,  {Iwasawa} K.,    {Johnstone} R.~M.,  2001, \mnras,
  321, L33

\bibitem[\protect\citeauthoryear{{Goldston} \& {Rutherford}}{{Goldston} \&
  {Rutherford}}{1995}]{1995Goldston}
{Goldston} R.~J.,  {Rutherford} P.~H.,  1995, Introduction to Plasma Physics.
Plasma Physics Series, Institute of Physics Pub.

\bibitem[\protect\citeauthoryear{{Hardcastle} \& {Croston}}{{Hardcastle} \&
  {Croston}}{2010}]{2010Hardcastle}
{Hardcastle} M.~J.,  {Croston} J.~H.,  2010, \mnras, 404, 2018

\bibitem[\protect\citeauthoryear{{Hardcastle}, {Kraft}, {Worrall}, {Croston},
  {Evans}, {Birkinshaw} \& {Murray}}{{Hardcastle}
  et~al.}{2007}]{2007Hardcastle}
{Hardcastle} M.~J.,  {Kraft} R.~P.,  {Worrall} D.~M.,  {Croston} J.~H.,
  {Evans} D.~A.,  {Birkinshaw} M.,    {Murray} S.~S.,  2007, \apj, 662, 166

\bibitem[\protect\citeauthoryear{{Hardcastle} \& {Krause}}{{Hardcastle} \&
  {Krause}}{2013}]{2013Hardcastle}
{Hardcastle} M.~J.,  {Krause} M.~G.~H.,  2013, \mnras, 430, 174

\bibitem[\protect\citeauthoryear{{Hardcastle} \& {Krause}}{{Hardcastle} \&
  {Krause}}{2014}]{2014Hardcastle}
{Hardcastle} M.~J.,  {Krause} M.~G.~H.,  2014, \mnras, 443, 1482

\bibitem[\protect\citeauthoryear{{Hogan}, {McNamara}, {Pulido}, {Nulsen},
  {Russell}, {Vantyghem}, {Edge} \& {Main}}{{Hogan} et~al.}{2017}]{2017Hogan}
{Hogan} M.~T.,  {McNamara} B.~R.,  {Pulido} F.,  {Nulsen} P.~E.~J.,  {Russell}
  H.~R.,  {Vantyghem} A.~N.,  {Edge} A.~C.,    {Main} R.~A.,  2017, \apj, 837,
  51

\bibitem[\protect\citeauthoryear{{Kalberla}, {Burton}, {Hartmann}, {Arnal},
  {Bajaja}, {Morras} \& {P{\"o}ppel}}{{Kalberla} et~al.}{2005}]{2005Kalberla}
{Kalberla} P.~M.~W.,  {Burton} W.~B.,  {Hartmann} D.,  {Arnal} E.~M.,  {Bajaja}
  E.,  {Morras} R.,    {P{\"o}ppel} W.~G.~L.,  2005, \aap, 440, 775

\bibitem[\protect\citeauthoryear{{Lazio}, {Cohen}, {Kassim}, {Perley},
  {Erickson}, {Carilli} \& {Crane}}{{Lazio} et~al.}{2006}]{2006Lazio}
{Lazio} T.~J.~W.,  {Cohen} A.~S.,  {Kassim} N.~E.,  {Perley} R.~A.,  {Erickson}
  W.~C.,  {Carilli} C.~L.,    {Crane} P.~C.,  2006, \apjl, 642, L33

\bibitem[\protect\citeauthoryear{{Mannering}, {Worrall} \&
  {Birkinshaw}}{{Mannering} et~al.}{2013}]{2013Mannering}
{Mannering} E.,  {Worrall} D.~M.,    {Birkinshaw} M.,  2013, \mnras, 431, 858

\bibitem[\protect\citeauthoryear{{Martel}, {Robichaud} \& {Barai}}{{Martel}
  et~al.}{2014}]{2014Martel}
{Martel} H.,  {Robichaud} F.,    {Barai} P.,  2014, \apj, 786, 79

\bibitem[\protect\citeauthoryear{{Mathews} \& {Guo}}{{Mathews} \&
  {Guo}}{2010}]{2010Mathews}
{Mathews} W.~G.,  {Guo} F.,  2010, \apj, 725, 1440

\bibitem[\protect\citeauthoryear{{Mathews} \& {Guo}}{{Mathews} \&
  {Guo}}{2012}]{2012Mathews}
{Mathews} W.~G.,  {Guo} F.,  2012, \apj, 755, 13

\bibitem[\protect\citeauthoryear{{McKean} et~al.,}{{McKean}
  et~al.}{2016}]{2016McKean}
{McKean} J.~P.,  et~al., 2016, \mnras, 463, 3143

\bibitem[\protect\citeauthoryear{{McNamara}, {Russell}, {Nulsen}, {Hogan},
  {Fabian}, {Pulido} \& {Edge}}{{McNamara} et~al.}{2016}]{2016McNamara}
{McNamara} B.~R.,  {Russell} H.~R.,  {Nulsen} P.~E.~J.,  {Hogan} M.~T.,
  {Fabian} A.~C.,  {Pulido} F.,    {Edge} A.~C.,  2016, \apj, 830, 79

\bibitem[\protect\citeauthoryear{{Nulsen}, {Young}, {Kraft}, {McNamara} \&
  {Wise}}{{Nulsen} et~al.}{2015}]{2015Nulsen}
{Nulsen} P.~E.~J.,  {Young} A.~J.,  {Kraft} R.~P.,  {McNamara} B.~R.,    {Wise}
  M.~W.,  2015, in {Massaro} F.,  {Cheung} C.~C.,  {Lopez} E.,
  {Siemiginowska} A.,  eds, Extragalactic Jets from Every Angle Vol.~313 of IAU
  Symposium, {Interaction of Cygnus A with its environment}.
pp 236--241

\bibitem[\protect\citeauthoryear{{Owen}, {Ledlow}, {Morrison} \& {Hill}}{{Owen}
  et~al.}{1997}]{1997Owen}
{Owen} F.~N.,  {Ledlow} M.~J.,  {Morrison} G.~E.,    {Hill} J.~M.,  1997,
  \apjl, 488, L15

\bibitem[\protect\citeauthoryear{{Panagoulia}, {Fabian} \&
  {Sanders}}{{Panagoulia} et~al.}{2014}]{2014Panagoulia}
{Panagoulia} E.~K.,  {Fabian} A.~C.,    {Sanders} J.~S.,  2014, \mnras, 438,
  2341

\bibitem[\protect\citeauthoryear{{Perley}, {Perley}, {Dhawan} \&
  {Carilli}}{{Perley} et~al.}{2017}]{2017Perley}
{Perley} D.~A.,  {Perley} R.~A.,  {Dhawan} V.,    {Carilli} C.~L.,  2017, \apj,
  841, 117

\bibitem[\protect\citeauthoryear{{Ponman}, {Cannon} \& {Navarro}}{{Ponman}
  et~al.}{1999}]{1999Ponman}
{Ponman} T.~J.,  {Cannon} D.~B.,    {Navarro} J.~F.,  1999, \nat, 397, 135

\bibitem[\protect\citeauthoryear{{Pratt}, {B{\"o}hringer}, {Croston}, {Arnaud},
  {Borgani}, {Finoguenov} \& {Temple}}{{Pratt} et~al.}{2007}]{2007Pratt}
{Pratt} G.~W.,  {B{\"o}hringer} H.,  {Croston} J.~H.,  {Arnaud} M.,  {Borgani}
  S.,  {Finoguenov} A.,    {Temple} R.~F.,  2007, \aap, 461, 71

\bibitem[\protect\citeauthoryear{{Reynolds}, {Lohfink}, {Ogle}, {Harrison},
  {Madsen}, {Fabian} et~al.,}{{Reynolds} et~al.}{2015}]{2015Reynolds}
{Reynolds} C.~S.,  {Lohfink} A.~M.,  {Ogle} P.~M.,  {Harrison} F.~A.,  {Madsen}
  K.~K.,  {Fabian} A.~C.,    et~al., 2015, \apj, 808, 154

\bibitem[\protect\citeauthoryear{{Sanders}}{{Sanders}}{2006}]{2006Sanders}
{Sanders} J.~S.,  2006, \mnras, 371, 829

\bibitem[\protect\citeauthoryear{{Smith}, {Wilson}, {Arnaud}, {Terashima} \&
  {Young}}{{Smith} et~al.}{2002}]{2002Smith}
{Smith} D.~A.,  {Wilson} A.~S.,  {Arnaud} K.~A.,  {Terashima} Y.,    {Young}
  A.~J.,  2002, \apj, 565, 195

\bibitem[\protect\citeauthoryear{{Smith}, {Brickhouse}, {Liedahl} \&
  {Raymond}}{{Smith} et~al.}{2001}]{2001Smith}
{Smith} R.~K.,  {Brickhouse} N.~S.,  {Liedahl} D.~A.,    {Raymond} J.~C.,
  2001, \apjl, 556, L91

\bibitem[\protect\citeauthoryear{{Snios} et~al.,}{{Snios}
  et~al.}{2018}]{2017Snios}
{Snios} B.,  et~al., 2018, \apj

\bibitem[\protect\citeauthoryear{{Steenbrugge}, {Blundell} \&
  {Duffy}}{{Steenbrugge} et~al.}{2008}]{2008Steenbrugge}
{Steenbrugge} K.~C.,  {Blundell} K.~M.,    {Duffy} P.,  2008, \mnras, 388, 1465

\bibitem[\protect\citeauthoryear{{Steenbrugge}, {Heywood} \&
  {Blundell}}{{Steenbrugge} et~al.}{2010}]{2010Steenbrugge}
{Steenbrugge} K.~C.,  {Heywood} I.,    {Blundell} K.~M.,  2010, \mnras, 401, 67

\bibitem[\protect\citeauthoryear{{Sutherland} \& {Bicknell}}{{Sutherland} \&
  {Bicknell}}{2007}]{2007Sutherland}
{Sutherland} R.~S.,  {Bicknell} G.~V.,  2007, \apjs, 173, 37

\bibitem[\protect\citeauthoryear{{Vantyghem} et~al.,}{{Vantyghem}
  et~al.}{2016}]{2016Vantyghem}
{Vantyghem} A.~N.,  et~al., 2016, \apj, 832, 148

\bibitem[\protect\citeauthoryear{{Vikhlinin}, {Kravtsov}, {Forman}, {Jones},
  {Markevitch}, {Murray} \& {Van Speybroeck}}{{Vikhlinin}
  et~al.}{2006}]{2006Vikhlinin}
{Vikhlinin} A.,  {Kravtsov} A.,  {Forman} W.,  {Jones} C.,  {Markevitch} M.,
  {Murray} S.~S.,    {Van Speybroeck} L.,  2006, \apj, 640, 691

\bibitem[\protect\citeauthoryear{{Wilman}, {Edge}, {Johnstone}, {Crawford} \&
  {Fabian}}{{Wilman} et~al.}{2000}]{2000Wilman}
{Wilman} R.~J.,  {Edge} A.~C.,  {Johnstone} R.~M.,  {Crawford} C.~S.,
  {Fabian} A.~C.,  2000, \mnras, 318, 1232

\bibitem[\protect\citeauthoryear{{Wilson}, {Smith} \& {Young}}{{Wilson}
  et~al.}{2006}]{2006Wilson}
{Wilson} A.~S.,  {Smith} D.~A.,    {Young} A.~J.,  2006, \apjl, 644, L9

\bibitem[\protect\citeauthoryear{{Wise}}{{Wise}}{in prep}]{2017Wise}
{Wise} M.~W.~o.,  in prep.

\bibitem[\protect\citeauthoryear{{Worrall} \& {Birkinshaw}}{{Worrall} \&
  {Birkinshaw}}{2006}]{2006Worrall}
{Worrall} D.~M.,  {Birkinshaw} M.,  2006, in {Alloin} D.,  ed., Physics of
  Active Galactic Nuclei at all Scales Vol.~693 of Lecture Notes in Physics,
  Berlin Springer Verlag, {Multiwavelength Evidence of the Physical Processes
  in Radio Jets}.
p.~39

\bibitem[\protect\citeauthoryear{{Worrall}, {Birkinshaw}, {Kraft} \&
  {Hardcastle}}{{Worrall} et~al.}{2007}]{2007Worrall}
{Worrall} D.~M.,  {Birkinshaw} M.,  {Kraft} R.~P.,    {Hardcastle} M.~J.,
  2007, \apjl, 658, L79

\bibitem[\protect\citeauthoryear{{Worrall}, {Birkinshaw}, {Young}, {Momtahan},
  {Fosbury}, {Morganti}, {Tadhunter} \& {Verdoes Kleijn}}{{Worrall}
  et~al.}{2012}]{2012Worrall}
{Worrall} D.~M.,  {Birkinshaw} M.,  {Young} A.~J.,  {Momtahan} K.,  {Fosbury}
  R.~A.~E.,  {Morganti} R.,  {Tadhunter} C.~N.,    {Verdoes Kleijn} G.,  2012,
  \mnras, 424, 1346

\bibitem[\protect\citeauthoryear{{Yaji}, {Tashiro}, {Isobe}, {Kino}, {Asada},
  {Nagai}, {Koyama} \& {Kusunose}}{{Yaji} et~al.}{2010}]{2010Yaji}
{Yaji} Y.,  {Tashiro} M.~S.,  {Isobe} N.,  {Kino} M.,  {Asada} K.,  {Nagai} H.,
   {Koyama} S.,    {Kusunose} M.,  2010, \apj, 714, 37

\bibitem[\protect\citeauthoryear{{Young}, {Wilson}, {Terashima}, {Arnaud} \&
  {Smith}}{{Young} et~al.}{2002}]{2002Young}
{Young} A.~J.,  {Wilson} A.~S.,  {Terashima} Y.,  {Arnaud} K.~A.,    {Smith}
  D.~A.,  2002, \apj, 564, 176

\end{thebibliography}

\end{document}